\documentclass{emulateapj}
\usepackage{graphicx}

\begin{document}

\title{Discovery, Photometry, and Kinematics
of Planetary Nebulae in M 82\altaffilmark{1,2}}

\author{L. C. Johnson\altaffilmark{3}, R. H. M\'endez\altaffilmark{4}, 
and A. M. Teodorescu\altaffilmark{4}}

\email{lcjohnso@astro.washington.edu}

\altaffiltext{1}{Based on data collected at the Subaru Telescope, which 
is operated by the National Astronomical Observatory of Japan.}

\altaffiltext{2}{This work was based on observations with the NASA/ESA 
{\it Hubble Space Telescope} obtained from the Multimission Archive at 
the Space Telescope Science Institute (MAST).  STScI is operated by
the Association of Universities for Research in Astronomy (AURA),
Inc., under NASA contract NAS5-26555.}

\altaffiltext{3}{Department of Astronomy, University of Washington, 
Box 351580, Seattle, WA 98195}

\altaffiltext{4}{Institute for Astronomy, University of Hawaii, 2680 
Woodlawn Drive, Honolulu, HI 96822}

\submitted{Accepted to ApJ: 23 March 2009}

\begin{abstract}
Using an [O {\textsc{iii}}] $\lambda$5007 on-band/off-band filter 
technique, we identify 109 planetary nebulae 
(PNe) candidates in the edge-on spiral galaxy M 82, using the 
FOCAS instrument at the 8.2m 
Subaru Telescope.  The use of ancillary high-resolution Hubble 
Space Telescope ACS H$\alpha$ imaging aided in confirming these 
candidates, helping to discriminate PNe from contaminants such 
as supernova remnants (SNRs) and compact H{\textsc{ii}} regions.  
Once identified, these PNe reveal a great deal about the 
host galaxy; our analysis covers kinematics, stellar 
distribution, and distance determination.  
Radial velocities were determined for 94 of these PNe
using a method of slitless spectroscopy, from which we obtain a clear 
picture of the galaxy's rotation.  Overall, our results agree with those 
derived by CO(2-1) and H{\textsc{i}} measurements (Sofue 1998) that 
show a falling, near-Keplerian rotation curve.  However, we find a 
subset of our PNe that appear to lie far ($\sim 1$ kpc) above the plane, 
yet these objects appear to be rotating as fast as objects close to the 
plane.  These objects will require further study to determine if they 
are members of a halo population, or if they can be interpreted
as a manifestation of a thickened disk as a consequence of a past
interaction with M 81.
In addition, [O {\textsc{iii}}] $\lambda$5007 emission line 
photometry of the PNe allows the construction of a planetary nebula 
luminosity function (PNLF) for the 
galaxy.  Our distance determination for M 82, deduced from 
the observed PNLF, yields a larger distance than those derived using 
the tip of the red giant branch technique (TRGB; Dalcanton et al. 2009), 
using Cepheid variable stars in nearby group member M 81 (Freedman 
et al. 1994), or using the PNLF of M 81 (Jacoby et al. 1989).  We show 
that this inconsistency most likely stems 
from our inability to completely correct for internal extinction imparted by 
this dusty, starburst galaxy.  Additional observations that yield 
object-by-object foreground and internal extinction corrections are
required to make an accurate distance measurement to this galaxy. 
\end{abstract}

\keywords{galaxies: individual (M 82) ---
          galaxies: kinematics and dynamics ---  
          planetary nebulae: general ---
          techniques: radial velocities}

\section{Introduction}

Studies of extragalactic planetary nebulae (PNe) have 
successfully contributed to astronomy in a number of 
different ways.  The use of the planetary nebulae 
luminosity function (PNLF) as a distance indicator 
(see e.g. Jacoby et al. 1989) has achieved remarkable 
success (Ciardullo 2003). Extragalactic PNe have 
been used to study galaxy cluster evolution and intracluster 
stellar populations in the Virgo cluster (Feldmeier et al. 
2004; Arnaboldi et al. 2008).  The fact that accurate radial velocities 
can be obtained from these PNe, once they are discovered, 
has made them increasingly important as kinematic probes in the outskirts
of galaxies. A classic example is the study of PNe in NGC 5128 by Hui
et al. (1995); they found evidence of dark matter around this 
elliptical galaxy, as expected from the currently favored scheme for 
galaxy formation. This finding put forth the
notion that further PN searches could lead to the accumulation of 
evidence supporting the ubiquity of dark matter in elliptical galaxies, 
as convincing as in the case of spiral galaxies.
However, discussion continues in the confusing case of 
``ordinary'' or ``average'' ellipticals, where PNe show little apparent 
influence from dark matter halos.  While the case for ``no dark matter'' is 
weakening as the effect of radial anisotropy on PN orbits is better
understood, recent studies by De Lorenzi et al. (2008), M\'endez et al. 
(2009), and Napolitano et al. (2009) show that clarification to this 
confusing situation will only come from further observations in elliptical
galaxies.  The fact is clear, however, that PNe act as versatile 
probes for studies of extragalactic astronomy.

In this paper we explore possible uses for PNe in the nearly edge-on,
starburst spiral galaxy M 82. To our knowledge no PN search in this 
galaxy has been tried up to now, probably because a large amount of dust
and internal extinction makes it very difficult to attempt a PNLF distance 
determination.

M 82 is relatively nearby (3.55 Mpc; Dalcanton et al. 2009),
making it a prime target for extragalactic PN identifications.
Its archetypal starburst activity and high 
infrared luminosity have induced many studies, 
giving us the added benefit of the availability of an assortment
of existing multi-wavelength observations.  This peculiar edge-on galaxy
has been found to have both a central bar (Telesco et al. 1991; Achtermann \&
Lacy 1995) and the signature of spiral arms (Mayya, Carrasco, \& Luna 2005).
Sofue et al. (1992) presented CO data that showed a rotation curve for the
galaxy that declined in nearly Keplerian fashion, making this system a 
unique example of a spiral galaxy that shows little apparent influence from 
dark matter.  Sofue (1998) goes on to suggest that the peculiar rotation shown 
in CO and H{\textsc{i}} data (Yun, Ho, \& Lo 1993) may be the result of an 
interaction with M 81 several hundred Myr ago.  Evidence of a galactic 
interaction exists in the form of a H{\textsc{i}} bridge and streamers connecting 
M 81, M 82, and NGC 3077 (Cottrell 1977; Yun, Ho, \& Lo 1994), and such
features were successfully reproduced in a numerical model of the encounter 
presented by Yun (1999).

\begin{figure*}[ht]
\epsscale{1.2}
\plotone{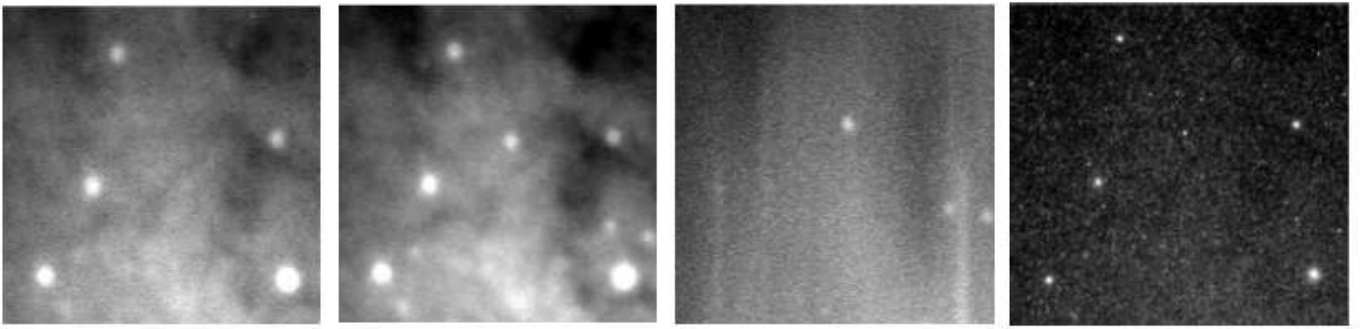}
\caption{Image subsections taken from final stacked images.  
From left to right: FOCAS
off-band, on-band, and grism (on-band + grism) image subsections;
at the extreme right, HST H$\alpha$ image subsection (see Section 
3.2).  Subsections shown are each $20''\!\! \times 20''\!\!$.  Three PN 
candidates are present in the on-band frame 
and can be identified by their absence in the off-band frame.  
The PN candidates appear again as point sources in the grism frame, 
shifted upwards in the direction of dispersion.  The displacement 
distance is a function of wavelength of the PN emission feature, 
which in turn provides the object's radial velocity. The high-resolution
HST image has superior ability to resolve non-PN objects and detect 
source extension. 
}
\label{fig_onoff}
\end{figure*}

Our initial focus was to test the accuracy of the slitless 
radial velocity method on a well-studied target and to confirm the 
ability of PNe to reproduce the behavior of a kinematically well-known 
population.  As spiral galaxies show distinctive kinematic qualities, such 
as a well-defined rotation curve, M 82 makes a good test object.
However, when we observed large numbers of PNe at great heights
from the galactic plane, the emphasis shifted toward exploring 
the distribution and kinematic behavior of the stellar population that 
is producing the detected PNe. We also wish to explore how reliable
a PNLF distance determination is in a difficult case like this.

In \S \ 2 we describe the technique for the detection of PNe, explain the 
method of slitless spectroscopy for determining radial velocities, and
introduce the PNLF. We discuss 
observations and image reductions in \S \ 3, followed by a consideration
of PN identifications and selection criteria in \S \ 4.
Radial velocity determination methods and results, as well as a discussion of
means for validating our findings, are presented in \S \ 5.  Photometry, PNLF
fitting, and our distance measurement are considered in \S \ 6.  
We close in \S \ 7 with a summary.

\section{Planetary Nebulae, Slitless Spectroscopy, and the PNLF}

The spectra of PNe characteristically show a very strong [O {\textsc{iii}}] 
$\lambda$5007 nebular emission line due to the 
excitation of a metastable level 
in a ground configuration by collisions with electrons. The colliding 
electrons are the result of ionization of low-density gas by extremely 
diluted UV photons from the PN central star. Due to the presence of this 
very strong [O {\textsc{iii}}] emission, PNe are easily detected using an 
on-band/off-band technique which involves blinking images taken through 
suitable on-band and off-band filters.  As shown in the first and second
frames of Figure \ref{fig_onoff}, while normal stars appear in both images, 
PNe can be identified by their differential appearance: absent in the 
off-band image, yet present in the on-band image.  Off-band images are exposed 
to a deeper limiting magnitude than the corresponding on-band image in 
order to ensure the validity of the non-detection.

An additional technique to substantiate the detection of a PN makes use 
of an on-band, grism-dispersed image.  This image is taken directly 
following the on-band image; the telescope continues 
tracking and the observing system is left unchanged between exposures, 
with exception to the addition of a dispersing grism placed 
into the light path.  In this configuration, the continuum light of a star 
will be shifted and dispersed into a segment of light on the image, whose 
length is determined by the bandpass of the filter used.  However, a PN 
will appear as a shifted point source, due to the nearly monochromatic
nature of its emission.  The third panel of Figure \ref{fig_onoff} clearly
shows that the three PN candidates detected by comparing the on-band and 
off-band images indeed appear as point sources on the grism image.

The power of the grism image not only lies in its ability to aid PN 
identification, but also in its ability to provide velocity data for 
each source.  The shift in position that occurs when we insert the grism 
is a function of the source emission's wavelength and its position in the 
field of view (due to the nature of the grism's 
dispersion and the distortions it creates). If we can 
successfully calibrate shifts as functions of wavelength and position on the 
CCD, we can calculate the wavelength of the PN emission, determine the 
redshift of the line, and thereby derive a radial velocity for the 
object using the Doppler redshift formula.

The calibration procedure consists of measuring the relationship between 
pixel shift and wavelength at specific locations across the entire CCD.  
If we make a sufficient number of these calibration measurements, an accurate 
interpolation of the desired relation at any intermediate location is 
possible.  Our calibration measurements are made at locations evenly placed 
at intervals of 100 pixels, creating a grid of calibration points covering 
the entire field of view.  Once a relation has been derived for each of these 
grid locations, a PN emission wavelength at any location within 
the grid may be 
calculated using the following procedure.  The four calibration 
positions closest to the undispersed image location of a PN are 
identified and the 
dispersed to undispersed shift of the PN is measured.  
This pixel shift is then 
converted to wavelength using the relations derived 
at each of the four surrounding 
calibration points, and bilinear interpolation is 
used to determine the final calibration
at the PN's undispersed pixel location and calculate 
a final emission wavelength value.

This method of PN radial velocity determination by means of slitless 
spectroscopy is an efficient and effective way to explore the kinematics 
of a galaxy.  Traditional methods of multiobject spectroscopy are high
overhead observations, where individual spectra must be taken
for each PN by means of complex multiple-slit masks or by
spectrographs equipped with optical fiber inputs.  Furthermore, this 
spectroscopic 
observation must be preceded by a separate imaging observation in order to
make the PN identifications.  The ability to obtain spectral information 
of multiple sources, typically in a single observing run, regardless of 
the number or distribution of objects in the field, makes slitless 
spectroscopy a valuable method of observation.

While the narrow-band imaging is initially used to make PN identifications 
and contributes to the velocity determination, much can be earned from the 
photometry of these objects as well.  The bright end of the PNLF has been 
shown to maintain a distinctive and consistent shape over a
wide range of galaxy types, luminosities, and distances, making it suitable
for use as a secondary standard candle for distance determinations out to
approximately 25 Mpc (Ciardullo 2003).  The PNLF is based on the
distribution of [O {\textsc{iii}}] $\lambda$5007 magnitudes in the PN
population of a galaxy.  We use a simulated PNLF similar to the one used
in M\'endez \& Soffner (1997) or 
in M\'endez et al. (2008) as our standard curve, to which we fit the
distribution of observed [O {\textsc{iii}}] magnitudes from our target 
galaxy. Our simulated PNLF simultaneously fits the total size of the 
population of PNe in the galaxy and the distance modulus.

\section{Observations and Image Reduction}

\subsection{Subaru+FOCAS Observations}

Observations of M 82 were made using the Faint Object Camera and 
Spectrograph (FOCAS, Kashikawa et al. 2002) at the Cassegrain focus 
of the 8.2m Subaru Telescope 
located atop Mauna Kea, Hawaii on three nights, 2004 November 6/7/8/9.  
FOCAS uses two 2k$\times$4k CCDs (15$\mu$m pixel size) covering a 
circular $7'$ diameter field of view, split by a $5''\!\!$ gap between
the CCD's.  The image scale is $0''\!\!.104$  
pixel$^{-1}$.  Observing conditions were photometric, with seeing 
better than or equal to $0''\!\!.7$.

Imaging was done using on-band and off-band filters with the 
following characteristics: effective central wavelengths of 5025 and 
5500 \AA, peak transmissions of 68\% and 95\%, and FWHM widths of 60 
and 1000 \AA.  The equivalent width of the on-band filter is 41.20 \AA \ 
for Chip 1 and 39.02 \AA \ for Chip 2.  The 
dispersed images were obtained using an Echelle grism with 175 grooves 
mm$^{-1}$, operating in fourth order, giving us a dispersion of 
0.5 \AA \ pixel$^{-1}$.  With 
$0.7''\!\!$ seeing, this translates into a resolution of 210 km s$^{-1}$.  

\begin{figure}[b]
\epsscale{1.3}
\plotone{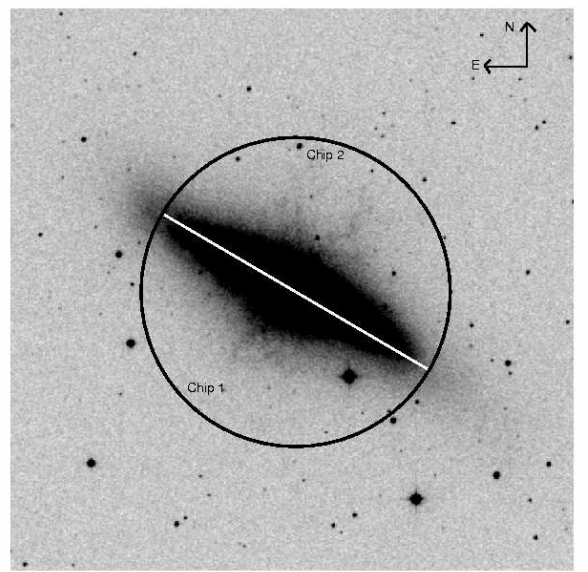}
\caption{Circular FOCAS field overlaid on DSS image centered on M 82. 
The white line approximately indicates the position of the 
narrow gap (5 arcsec) separating Chip 1 from Chip 2.
}
\label{field}
\end{figure}

Figure \ref{field} shows the observed field.
The images can be divided into the following groups:

M 82 Images:\\
1. Off-band images of M 82.\\
2. On-band images of M 82.\\
3. On-band plus grism images of M 82.

Photometric Calibration Images:\\
1. On-band images of LTT 9491 (on each chip).

Slitless Spectroscopy Calibration Images:\\
1. On-band images of engineering mask illuminated by Th-Ar calibration 
lamp to acquire undispersed calibration positions.\\
2. On-band plus grism images of engineering mask illuminated by Th-Ar 
calibration lamp to acquire dispersed calibration spectra.

Slitless Spectroscopy Calibration Control Images:\\
1. On-band images of engineering mask illuminated by the large, local PN
NGC 7293, to acquire undispersed source positions.\\
2. On-band plus grism images of engineering mask illuminated by NGC 7293 to 
acquire dispersed source spectra.\\
3. On-band plus grism images of engineering mask illuminated by Th-Ar 
calibration lamp to acquire dispersed calibration spectra.

Our data consist of six sets of observations (two on each night), 
where each set is comprised of one image from each of 
the three M 82 image types 
and the two slitless spectroscopy calibration image groups.  
In addition, nightly sets of photometric calibration images and 
slitless spectroscopy
calibration control images.  The use of these calibration control 
images will be 
fully described in \S \ 5.  All images underwent basic CCD reductions 
(bias subtraction and flat-field correction using sky 
flats) using IRAF\footnote[2]{IRAF is distributed by the National Optical
Astronomical Observatories, which are operated by the Association of
Universities for Research in Astronomy, Inc., under cooperative agreement
with the National Science Foundation.} standard tasks.

Table \ref{tbl-1} lists the most important 
CCD images obtained for this project.
Since there are two CCD images for each field (Chips 1 and 2),
for brevity we have listed only the exposure number corresponding to
Chip 1. For Chip 2, add 1 to the listed exposure number.

The six sets of on-band, off-band, and grism images 
were registered and combined for each of the 
two CCD chips.  The IRAF procedure \texttt{geomap} was 
used to calculate appropriate registration shifts and 
\texttt{imshift} was used to perform these 
shifts.  Registration errors for the on-band 
and off-band images were smaller than 
0.2 pixels, however errors for the grism 
images were greater.  Grism registration 
errors were typically 0.3 pixels, with one 
frame having an error of 0.5 pixels.  Sets of
registered images were combined and cosmic rays
removed using \texttt{imcombine} and its cleaning 
algorithm  \texttt{minmax}.  The off-band combined image 
was then registered to the combined on-band image 
for PN identification purposes, 
while the combined grism image needed no 
registration to the combined on-band image, as it was imaged in an 
identical instrument position as the on-band. 

When the observations were made, 
the field of view was oriented so that the 
major axis of M 82 would fall near 
the $5''$   gap between the two CCDs.  
This choice was made due to the expectation that very few PNe would be
visible along the major axis of the galaxy because of 
heavy dust extinction and bright, confusing background continuum 
emission along the extent of the disk.  For analysis and cataloging
purposes, the major axis was defined by fitting 
to peaks in the galaxy continuum emission.  Emission peaks
were identified fitting a high-order polynomial to 
image cuts made in the minor axis 
direction through regions of smooth continuum emission 
located at the ends of the 
visible disk.  Once the major axis was defined, 
we determined the minor axis to be the 
line perpendicular to the major axis that passed 
through the kinematic center of the 
galaxy.  We adopted the kinematic center defined 
by Achtermann \& Lacy (1995) at
$\alpha$= $9^{\rm h}55^{\rm m}52^{\rm s}.14$ 
$\delta$=$69^{\circ}40^{'}45''\!\!.90$, 
thus placing our major/minor axis coordinate center at 
$\alpha$= $9^{\rm h}55^{\rm m}51^{\rm s}.77$ 
$\delta$=$69^{\circ}40^{'}50''\!\!.22$.  These 
axes are rotated $4^\circ$ from the original pixel axes.

\subsection{HST+ACS Observations}

To complete our set of observations, we used a Hubble Space Telescope
six-point mosaic H$\alpha$ image of the galaxy, observed using the 
ACS wide-field imager.  These observations were made available via the 
Multimission Archive at STScI.  For details on the observation and image
reduction, please see Mutchler et al. (2007), but we include a short summary
here.  Twenty-four individual overlapping, dithered exposures from the 
ACS Wide Field Channel, consisting of two 2k$\times$4k pixel CCD chips,
were combined to create a single 12k$\times$12k pixel (10.24$\times$10.24
arcmin) image.  The image scale is $0''\!\!.05$ pixel$^{-1}$.  
Imaging was made through a F658N H$\alpha$ filter, with the following 
characteristics: effective central wavelength of 6584  \AA, peak 
transmission of 44\%, and FWHM width of 87 \AA. The PSF 
FWHM is $\sim 0''\!\!.1$.  The right-most panel
of Figure \ref{fig_onoff} shows a cutout from the ACS H$\alpha$ image,
to illustrate its excellent resolution in comparison to the FOCAS 
cutouts.  

The mosaiced H$\alpha$ image has a total exposure time of 3320 s.  
When combined with a superior observing platform and space-based 
resolution, this image depth is more than adequately matched to our 
[O\textsc{iii}] imaging.  All candidate PN objects 
identified in the [O\textsc{iii}] imaging were detected (S/N $>$ 3) in the 
H$\alpha$ image.  

\section{PN Identification}

In \S \ 2 we discussed using an on-band/off-band 
blinking technique for making PN identifications.  The first step in
applying this method was to produce additional images to aid our search.  
A version of each of the FOCAS combined images was created by 
using the IRAF \texttt{fmedian} filtering task to subtract the smooth
background galaxy continuum emission.  These resulting images
proved to be essential in making PN identification within regions 
close to the major axis of the galaxy that before
had bright, confusing backgrounds.  In addition, an image
was created using analysis software developed by Alard \& Lupton (1998)
and implemented in Munich by G\"ossl \& Riffeser (2002) that
carefully subtracts a properly scaled off-band image 
from an on-band image.  The result is a high quality
continuum subtracted image, where only desired emission line sources remain.  
This image not only aided in identification purposes, but will play an
important role in our photometry work, as detailed in \S \ 6.

We visually inspected our data using three initial selection criteria: 
detection in the on-band image, non-detection in the off-band image, 
and point source detection in the grism image.  The result of this initial 
effort was the identification of 116 preliminary PN candidates.  To obtain 
accurate coordinates for each of the candidates, we created a coordinate 
transformation from pixel coordinate positions on the FOCAS images to 
pixel coordinate positions on the ACS images using IRAF tasks
\texttt{geomap} and \texttt{geoxytran}.  
Source identifications on the ACS images were 
confirmed by eye and accurate centers were calculated using 
\texttt{imcentroid}.  Final coordinates were obtained using \texttt{skyctran} 
from the WCS coordinates embedded in the ACS image header.  These 
WCS coordinates are good to 0.1 pixels, however the final accuracy of 
our positions are limited to $0''\!\!.05$ by the uncertainty of 
our centering results.

In our search for PNe in a late-type galaxy 
such as M 82, the risk of contamination from numerous 
supernova remnants (SNRs) and H{\textsc{ii}} regions in a 
galaxy with active on-going star formation is great.  To combat
this issue, we imposed two additional criteria for the confirmation
of our PNe candidates: they appear as unresolved point sources 
with no visible extension and they show a characteristically high 
line ratio of $ R=I([O\textsc{iii}]) / I(H\alpha) $.

The criterion concerning the rejection of resolved sources or those showing
visible extension is included to differentiate between true PNe and
contaminants using a candidate's angular size.  Bright galactic PNe
have sizes less than $\sim 1$ pc (Acker et al. 1992).  At M 82's 
distance of $\sim 3.5$ Mpc, 
this corresponds to an angular size for PNe of less than $0''\!\!.06$, which 
cannot be resolved in even our highest resolution H$\alpha$ image.  
Therefore, if we observe any resolved sources or extended structures in 
this image, these objects must have extents on the order of tens of parsecs, 
thus eliminating them as potential PNe.  As further proof of this point, we 
conducted a search for known SNRs and H{\textsc{ii}} regions.  
Optical and radio surveys of these objects have mainly focused on 
the starbursting core, within 1 kpc of the nucleus (de Grijs, et al. 2000; 
Rodriguez-Rico et al. 2004).  While this 
was not a favored region for PN discovery, due to the bright, confusing
background, none of our PN candidates were coincident with these objects.
Furthermore, these studies derived sizes of these objects that were all 
greater than $\sim 2$ pc, making them at least marginally resolved 
objects in our ACS imaging.

The final selection criterion for PN identification uses
[O{\textsc{iii}}] and H$\alpha$ data to calculate 
a line ratio between the two:
\begin{equation}
R=I([O\textsc{iii}]) / I(H\alpha + [N\textsc{ii}]).
\end{equation}
For simplicity, we will refer to the H$\alpha$ + [N{\textsc{ii}}] blend 
as simply H$\alpha$ emission throughout. 
While the previous criterion on resolved sources should eliminate 
most contaminants, there is still a possibility 
that compact H{\textsc{ii}} regions could still remain
in our sample.  An important distinguishing feature 
of PNe, however, is that the central stars responsible for 
the ionization of PNe are typically higher 
temperature objects than the stars that ionize H{\textsc{ii}} regions.  
Because of this fact, we expect the levels of 
doubly ionized oxygen emission, with respect to H$\alpha$ emission, 
to be higher in PNe than in H{\textsc{ii}} regions.  A survey by 
Shaver et al. (1983) showed that most H{\textsc{ii}} regions 
$(\geq 80 \%)$ have $R < 1$.  Therefore, we can assume 
that the line ratio, R, will in general be greater than one for PNe, 
while the opposite is true for H{\textsc{ii}} regions.

\begin{figure}
\epsscale{1.25}
\plotone{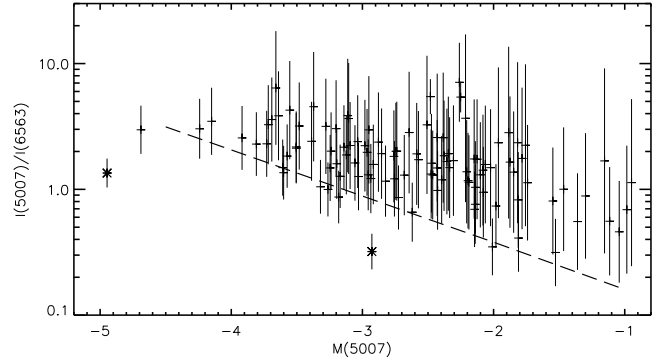}
\caption{[O{\textsc{iii}}]/H$\alpha$ ratio 
plotted as a function of the absolute
magnitude $M$(5007), as defined in Section 5.1. The dashed 
diagonal line is the PNe-H{\textsc{ii}} boundary as defined by Herrmann 
et al. (2008) using the Dalcanton et al. (2009) TRGB distance to calculate
$M$(5007). Sources located below the dashed 
diagonal line should be rejected as possible H{\textsc{ii}} regions, therefore 
the two candidates marked with asterisks were removed from the sample.
}
\label{Herrmann}
\end{figure}

Herrmann et al. (2008) expand on this idea and use
observations presented in Figure 2 of Ciardullo et al. (2002) to show 
the tendency of PNe to populate a distinctive region 
in [O{\textsc{iii}}] - H$\alpha$ emission line space. Using this fact, they 
define a limiting R as a function of absolute [O{\textsc{iii}}] magnitude in their 
Eqs. 2 \& 3, below which PN candidates should be rejected.  This boundary 
sets a stringent standard for the minimum allowable R value 
over the top $\sim1.5$ mag of the PNLF, whose members are 
critical for making a PNLF distance determination.
Figure \ref{Herrmann} shows the positions of our M 82 sources 
in the Herrmann et al. (2008) plot.  It should be noted that as this boundary
is set relative to the absolute [O{\textsc{iii}}] magnitude, distance
is intrinsically incorporated into this plot.  In addition, because we can 
assume that there will be some extinction from such a dusty, 
edge-on galaxy, we must also remind the reader that this plot is
subject to extinction effects.  These include extinction in 
[O{\textsc{iii}}] that will tend to move objects to fainter absolute 
[O{\textsc{iii}}] magnitudes and differential extinction that will lower
R values.

Of the 116 original candidates that satisfied 
the initial three criteria, we eliminated four 
due to suspicious extension and resolved sizes, 
one due to a low [O{\textsc{iii}}]/H$\alpha$
ratio, and one that was disqualified by both of the
additional criteria. Approximate derived sizes 
for the resolved objects ranged from $\sim15 - 350$ pc.  Even
when allowing for up to $\sim 0.5$ mag of uncertainty in the x-axis
location of objects in Figure \ref{Herrmann}, two objects
clearly lie outside the accepted range of R values.  One additional 
candidate was also removed due to an anomalous velocity measurement, 
which we will discuss more throughly in \S \ 5.  The
seven rejected candidates, their positions, and 
other derived data can be found in Table \ref{mybadtable}.  The remaining
109 objects that make up our final sample, their positions, 
and all other derived data, can be found in Table \ref{mytable}.  

\section{Radial Velocity: Calibration and Results}

\subsection{Calibration and Velocity Determination}

We introduced the procedure to obtain slitless radial velocity measurements
for our PN candidates in \S \ 2.  We will now describe the specific details 
behind the calibration
procedure.  The layout of points that sets up our calibration grid, for
later use in bilinear interpolation, is created by placing an 
engineering mask, containing 970 holes in the desired 100 pixel interval grid 
pattern, into the light path.  Illuminating the CCD with a Thorium-Argon 
calibration lamp through both the mask and the on-band filter defines 
the undispersed positions for each of the calibration points.  After 
inserting the grism, the light from the calibration lamps is dispersed 
into the Th-Ar spectrum, from which the positions of five Th and Ar emission 
lines can be measured.  The shifts from the undispersed to the dispersed 
positions for each of these known wavelengths are calculated, and the relation
between shift and wavelength is determined, allowing for eventual 
wavelength interpolation for our PNe candidates.  The accuracy of this 
calibration technique is explored in \S \ 5.2.

\begin{figure}[!h]
\epsscale{1.2}
\plotone{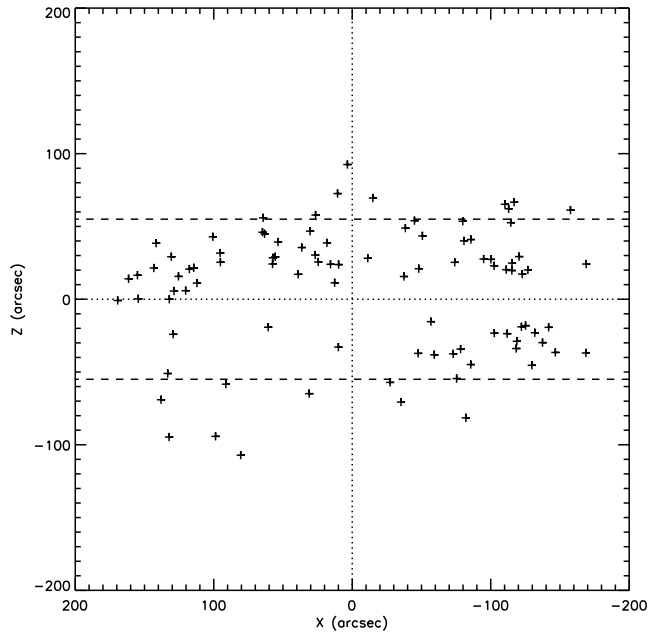}
\caption{Positions of the 94-member PN velocity sample, relative to 
the center of M 82.  Coordinates $x$ and $z$ are measured along the 
major and minor axes, respectively.  Dashed lines denote $z$ distance
from the major axis of 55 arcsec, equivalent to $\sim 0.95$ kpc at the 
distance of M 82.
}
\label{position}
\end{figure}

Each candidate's ($x$,$y$) pixel position on the median versions of the 
non-dispersed on-band image and the dispersed grism image were obtained using 
IRAF's \texttt{imcentroid} task, and dispersion shifts were calculated 
for each source.  After calibrating our data as described 
above, using a set of calibration images taken during the first night of
observations, we used a custom bilinear interpolation code 
to calculate wavelengths
and velocities for the PN candidates.  Fifteen objects 
were left unmeasured due to their location on the image outside of the 
calibration grid, making complete, accurate interpolation impossible.  
In addition, one PN candidate was removed from the sample due to its extremely 
discrepant velocity of -251 km s$^{-1}$, as referred to in \S \ 4.  We assume 
this object is a background emission-line galaxy 
that has a redshifted emission line within the bandpass of the filter, thus 
explaining the single inconsistent measurement.  After combining this single 
rejection with those already detailed in \S \ 4, 
and considering those objects left 
unmeasured, we come to our final 
sample of 94 confirmed PNe with measured radial 
velocities.  This velocity data, along with the 
previously mentioned positional 
data, are included in Table \ref{mytable}.  
The uncertainty in the velocities is 
10 km s$^{-1}$, based upon the grism dispersion, seeing conditions, 
and uncertainty in image registration.
The positions of these final velocity 
sample members, relative to the center of M 82,  
are shown in Figure \ref{position}. The $x$ and $z$ coordinates are defined, 
respectively, along the major and minor axis.

\subsection{Calibration and Slitless Velocity Analysis}

\begin{figure}
\epsscale{1.15}
\plottwo{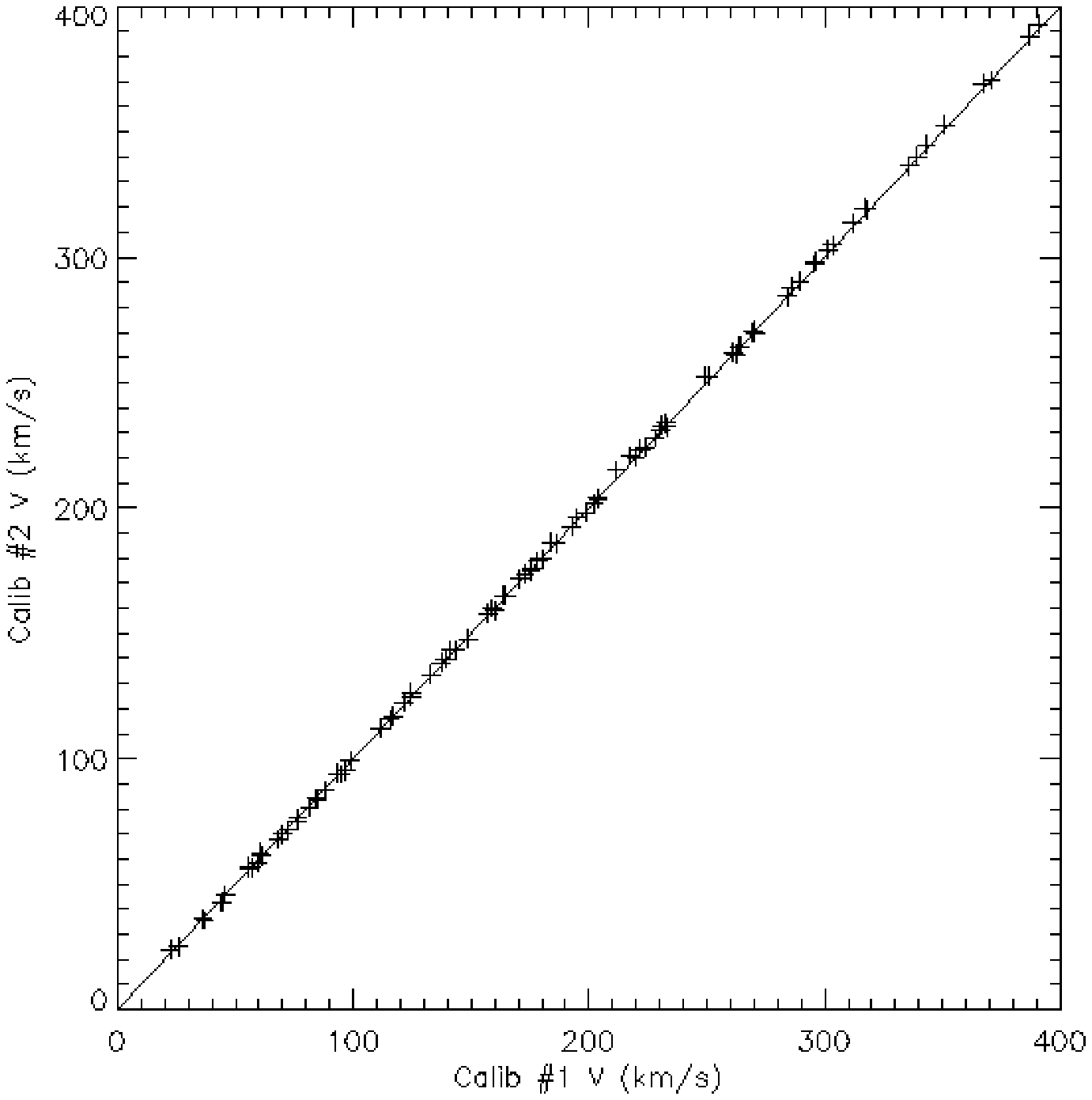}{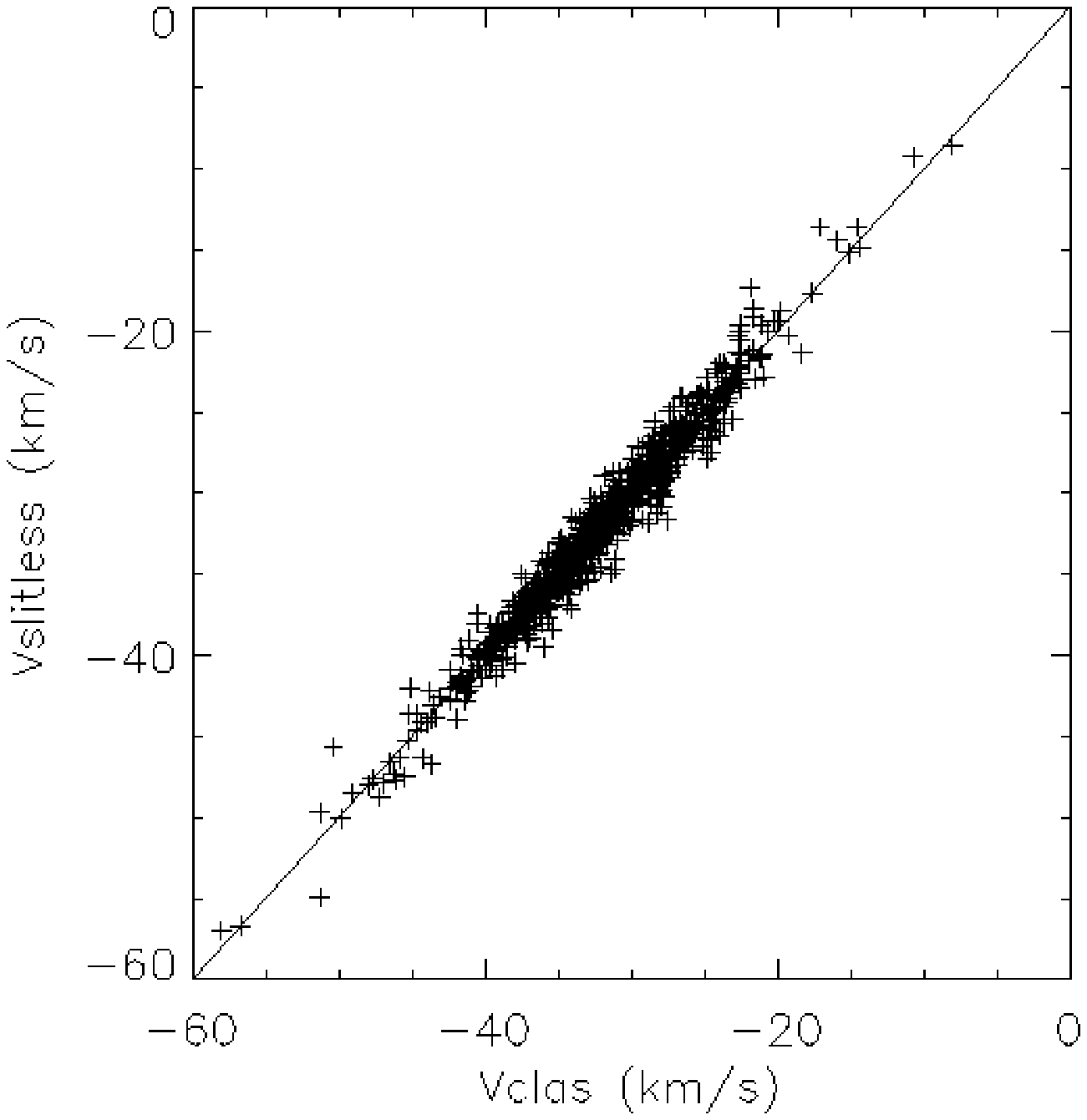}
\caption{Calibration analysis results.  Left: Comparison of radial 
velocity measurements made for M 82 PN sample using two different 
sets of calibration images.  Right: Comparison of radial velocity 
measurements made for NGC 7293 using classical 
and slitless techniques for velocity determination.
}
\label{calibdiff}
\end{figure}

To confirm the accuracy of the slitless velocity measurements, we 
conducted two forms of analysis on our calibrations.  The first analysis 
technique was simply to use a second set of calibration images to 
calibrate our velocities.  The left panel of Figure \ref{calibdiff} 
shows that the velocities determined using the different calibration 
files are virtually identical.  The maximum difference between 
measurements was 3 km s$^{-1}$, and the mean 
difference was 1 km s$^{-1}$.  The second portion of the
analysis was to compare our technique of slitless velocity determination 
to that of a classical (slit) method.  The local Galactic PN NGC 7293 
was chosen as a test case.  Images of NGC 7293 were taken in 
two different ways: an on-band image acquired 
through the engineering mask, with 970 holes covering the 
extent of the field
of view, and a dispersed on-band image acquired through 
both the grism and the engineering mask.  The angular 
size of NGC 7293 is such that it fills the 
entire field of view, so that by inserting the 
engineering mask, we create 970 point sources from our 
single target object.  
This provides us with an opportunity to measure 
velocities, at locations spread across the entire 
field of view, that should agree, on average, with the 
known velocity of NGC 7293. Some scatter is expected, 
due to variations in the PN's expansion velocity field and density 
distribution that each contribute to dispersion in individual 
velocities measured at points across the field of view. 
We can make the radial velocity measurements in the same 
manner as for M 82, using our slitless method, or we can treat each 
hole as a slit and measure the velocity of 
the source in the classical way, by making 
a direct comparison of the dispersed location 
of the target emission line with the 
locations of the calibration spectral features.  
The right panel of Figure \ref{calibdiff} shows that the velocities 
determined using the two different methods 
agree very well with each other.  In addition, 
the average of 916 radial velocity measurements we obtained for 
NGC 7293 is $-32$ km s$^{-1}$, which shows 
close agreement with the known value of $-27$ 
km s$^{-1}$ (Meaburn et al. 2005).  The
scatter we find, from $-60$ to $-5$ km s$^{-1}$, is
also in close agreement with the findings of other comparable 
observations of the NGC 7293 (see Fig. 8, Meaburn et al. 2005). 
From all these facts, we conclude that the uncertainty in our 
velocity measurements is at most 10 km s$^{-1}$.

There is a final point to consider: the airmass of our NGC 7293 validation
observations is smaller than that of our M 82 observations
(see Table 1), and possible flexure in the FOCAS spectrograph might
produce systematic velocity differences between these two telescope
alignments.  We can show that this is not the case by examining our
previous study of PNe in NGC 4697 (M\'endez et al. 2009). The observations 
of NGC 7293 used in that work were obtained at an airmass of about 1.8,
which is similar to the airmass of our M 82 observations.
Inspection of Section 4 and Figure 3 in M\'endez et al. (2009)
shows that there is no significant difference in the radial
velocities measured across NGC 7293 with respect to what we find in
our analysis, displayed in Figure \ref{calibdiff}. In other words, the mechanical 
stability of FOCAS is such that flexure effects are negligible for our purposes.

\subsection{Velocity Results: Comparison With Existing Literature}

\begin{figure}
\epsscale{1.15}
\plotone{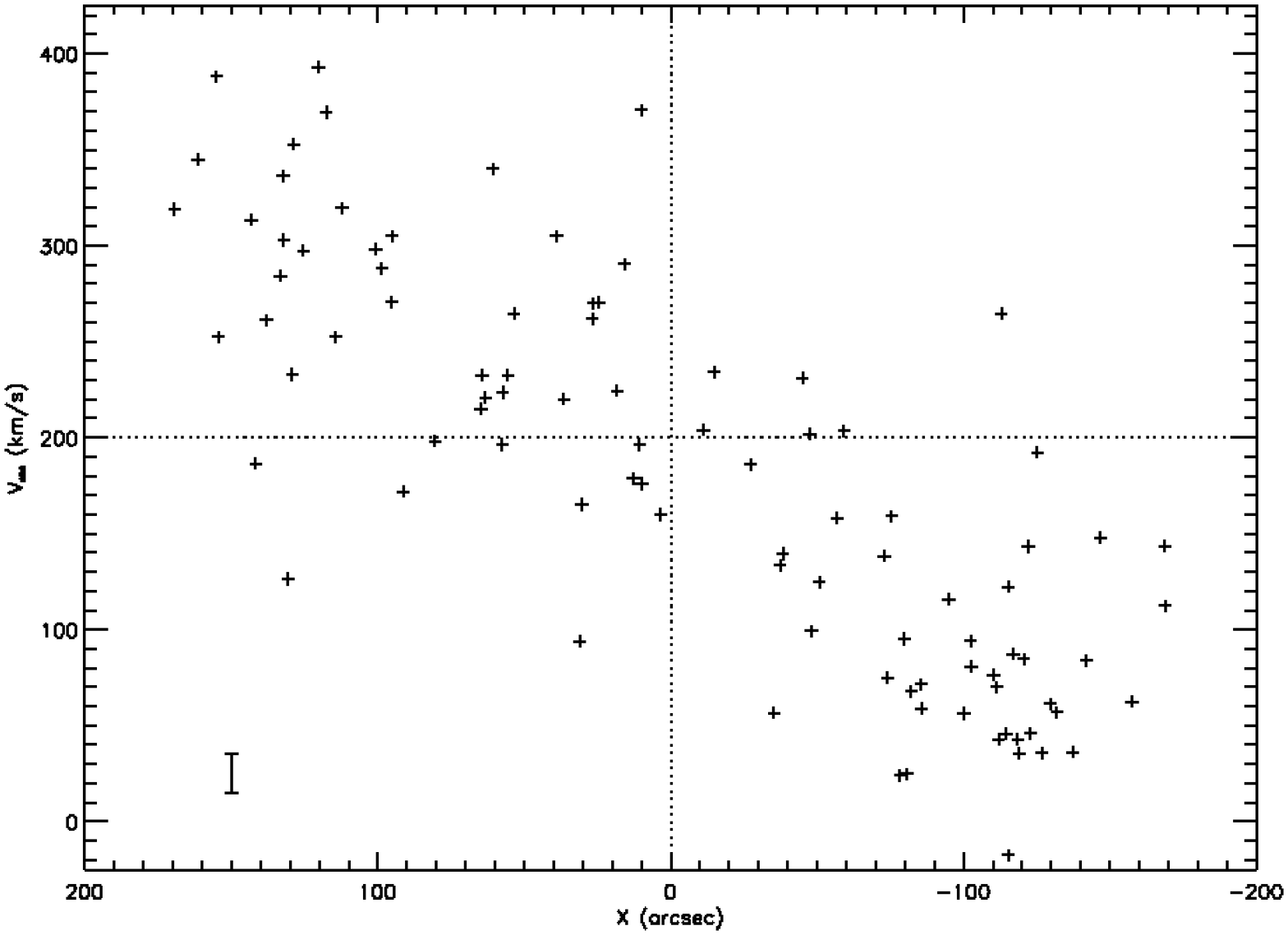}
\caption{Radial velocities of the complete velocity sample, numbering
94 PNe, plotted as a function of their major axis position.  The error bar
located in the bottom left is representative of the 10 km s$^{-1}$ 
uncertainty for each data point.}
\label{xvsvel}
\end{figure}

Our velocity measurements of PNe in M 82
show the expected signature of a rotating disk.  
Figure \ref{xvsvel} shows 
that the northeast, receding side of the disk 
(positive offset) is redshifted to 
velocities greater than that of the systemic velocity, 
while the southwest, approaching side of the disk 
(negative offset) is blueshifted to 
velocities less than that of the systemic velocity.  This
signature of the bulk rotation of the galaxy agrees with published M 82
velocity data (e.g. Sofue et al. 1992) and the trailing direction of its
spiral arms (Mayya et al. 2005).  

We also find agreement with values of M 82's systemic velocity.  
Achtermann \& Lacy (1995) and McKeith et al. (1993) both derived 
a systemic velocity for M 82 of
200 km s$^{-1}$ with uncertainties of 7 and 11 km s$^{-1}$ respectively.  
When we narrow our sample to 43 near-disk objects 
found within $\sim 0.5$ kpc ($29''\!\!$ ) 
of the major axis, to reduce the dilution of the rotational signature, we 
measure a systemic velocity of $195 \pm 18$ km s$^{-1}$.  This agreement with the
literature value, within the range of uncertainty, provides 
further evidence of the accuracy of our data.  In subsequent analysis 
for this study, we adopt the literature value of 200 km s$^{-1}$ for the
systemic velocity, due to our measurement's limited sample size.

Rotation curves are a standard tool 
central to the characterization of the kinematic 
properties of a galaxy.  Sofue et al. (1992) 
presents a rotation curve for M 82, derived
from CO(2-1) observations, that shows a peculiar, 
near-Keplerian decline, to which
we would like to compare our results.
Due to the uncertainty in the height from the 
galaxy mid-plane of our observed PNe,
projecting the positions of these
PNe onto the plane in order to plot a comparable rotation curve becomes 
problematic.  Therefore, we choose instead to 
compare our results to a simulated 
observational signature we would expect from a PN population that follows
the CO-derived rotation curve.

These simulations model the spatial and 
kinematic distributions of an observed PN
population, which should help in the 
interpretation our velocity results.  Spatially, we 
assume that PNe will be distributed 
radially in an exponential disk.  For this exponential,
we choose the K-band scale length, as 
derived by Mayya et al. (2005).  For simplicity, 
we neglect adding a vertical scale height distribution 
and forgo modeling structural features 
such as spiral arms.  To help fit our 
observations, we also include a second extincting
exponential distribution, with a shorter 
scale length, which results in a final ring-like 
particle distribution that  mimics the 
extinction towards the center of the galaxy that
prevents any detections from this region.  
Particles are then randomly distributed
according to this distribution and assigned 
velocities according to either a typical
flat rotation curve or a falling rotation 
curve that matches the CO data.  These velocities
also take into account a velocity dispersion 
of 60 km s$^{-1}$, as is appropriate for a 
system with a peak rotational velocity 
of $\sim 200$ km s$^{-1}$ (Bottema 1993).  The final step
is to then take this simulated disk galaxy 
and transform its inclination and velocities
to conform to our observed view, adopting 
an inclination of 77 degrees and a systemic
velocity of 200 km s$^{-1}$ (Mayya et al. 2005; McKeith et al. 1993)

\begin{figure*}
\epsscale{1.15}
\plotone{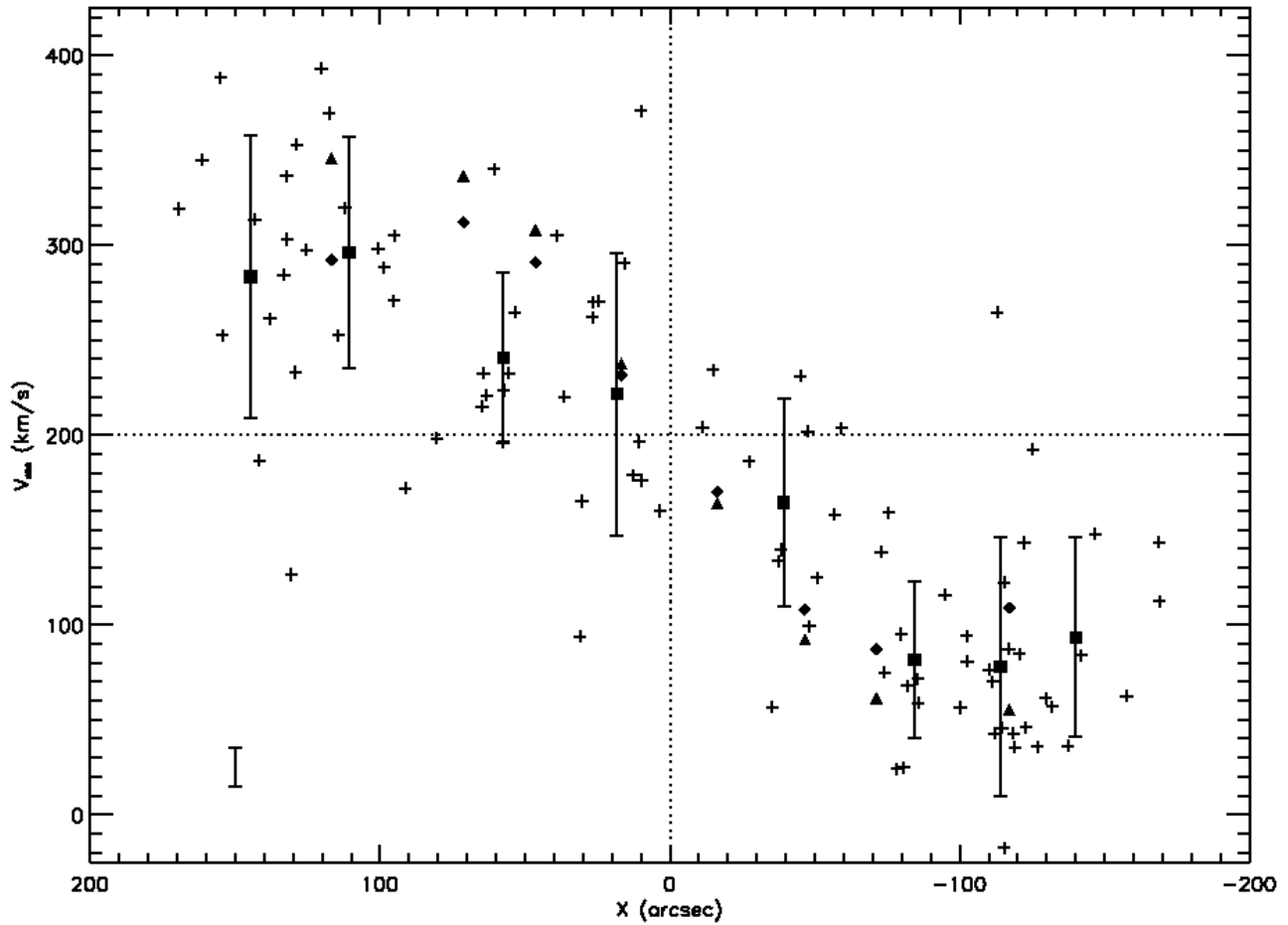}
\caption{Binned velocities of 94 PNe as a function of their major axis 
position.  Our bins are represented by square symbols, where the 
numbers of PN in each bin, from left to 
right, are 11, 12, 11, 12, 12, 12, 12, 12.  
Bars on squares represent the 
dispersion within the binned velocities, 
in this case ranging from 41 to 75 km s$^{-1}$.  
Diamond symbols represent binned simulation results for a falling 
input rotation curve, while triangle symbols represent binned
simulation results for a flat input rotation curve.  Individual 
PN velocity data points, showing the radial velocity distribution 
of our sample, are denoted with plus signs.  The error bar
located in the bottom left is representative of the 10 km s$^{-1}$ 
uncertainty of each individual data point.}
\label{binvel}
\end{figure*}

As we run each simulation, we bin the resulting 
velocity distributions into eight bins, according 
to their position along the observed 
major axis.  By then binning our own data 
in the same manner, it is possible to see
how our velocity distribution compares with 
that of our simulations.  These binned
results are plotted in Figure \ref{binvel}.  
This figure shows that we find good
agreement between our binned velocities 
and the simulated falling rotation curve
at positive $x$ positions.  However, at 
negative $x$ positions, we find that our binned
averages seem to fall directly between the 
simulated flat and falling rotation curve 
data points.  One possible source of this 
discrepancy lies in our approximation 
of a featureless exponential disk.  It 
appears that there is an overdensity of individual 
PN velocity data points in Figure \ref{binvel} centered at approximately 
($X$,$V$)=($-110$, 40).  Interestingly, this 
grouping also appears in position space in
Figure \ref{position} at ($X$,$Z$)=($-110$, 25).  
One possibility for these groupings
could be that these structures are part of 
the approaching, western spiral arm, as
identified in Mayya et al. (2005).  This would 
cause our observed velocities to differ
from our idealized simulation due to a possible 
bias in the average binned velocity
of our data.

While it is possible, with a few caveats, to 
find agreement between our velocity data 
and simulations of a declining, near-Keplerian 
rotation curve, we cannot be 
completely sure with regard to the shape of the 
rotation curve of our PN population.
However, the multiple pieces of evidence presented 
in this section certainly favor
agreement between the kinematic picture depicted by our PN velocities and 
published data found in the literature.

\subsection{Velocity Results: PNe at High $z$}

It is immediately apparent in Figure \ref{position} 
that we have identified many 
PNe at rather large 
values of $z$. Given the inclination of approximately 77 degrees (Mayya
et al. 2005), this indicates that these objects are likely to 
lie at large heights above 
the galactic plane.
Assuming a distance to M 82 of 3.55 Mpc, PNe at $z > 55$ arcsec are located
$\sim 0.95$ kpc from the plane when projected from the major axis.  

\begin{figure*}
\epsscale{1.15}
\plotone{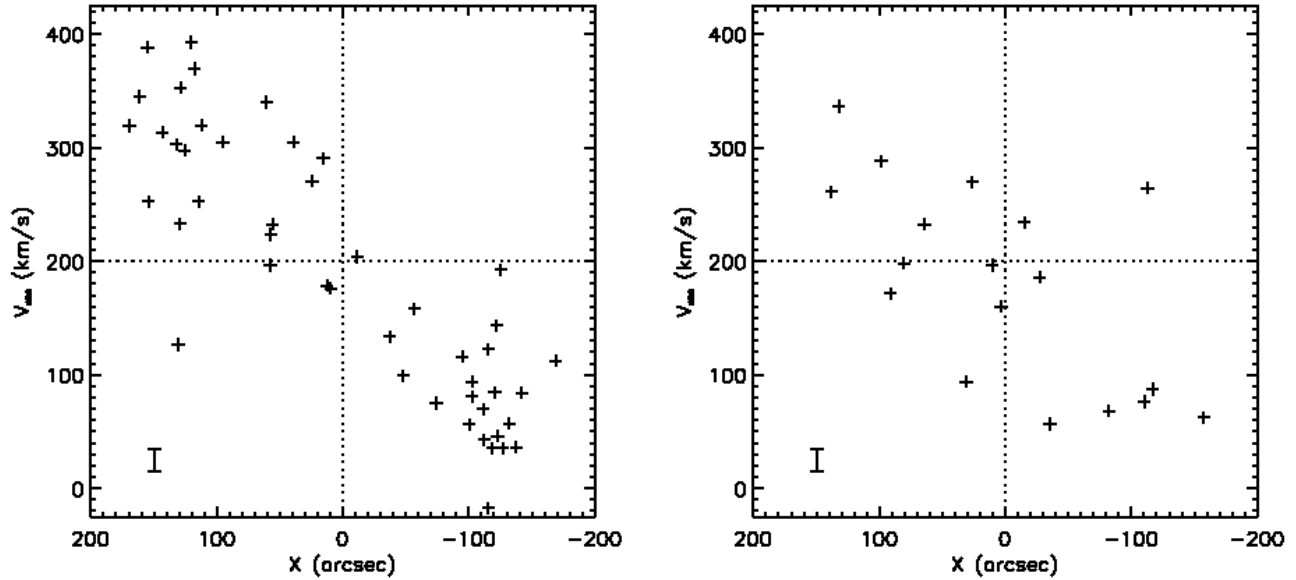}
\caption{Left: PN radial velocity as a function of $x$ coordinate, for 
PNe with $z < 30$ arcsec. Right: Same, for PNe with $z > 55$ arcsec. 
The systemic velocity of M 82 is 200 km s$^{-1}$, and the error bar
located in the bottom left of each plot is representative of the 10 km s$^{-1}$ 
uncertainty for each data point.
}
\label{halokin}
\end{figure*}

Figure \ref{halokin} shows a comparison of the rotation shown by two
PN groups in M 82: those PNe with $z < 30$ arcsec are displayed in the 
left plot, while those with $z > 55$ arcsec are displayed in the right plot. 
The substantial rotation at high $z$, comparable to the rotation found 
near the major axis, is remarkable.  This result conflicts with the
findings of Sofue et al. (1992) who, using CO velocities, found slow
rotation speeds in the range of 20-30 km s$^{-1}$ at $z \sim 1$ kpc, and 
little or no rotation at $z > 1.4$ kpc.

\begin{figure}
\epsscale{1.25}
\plotone{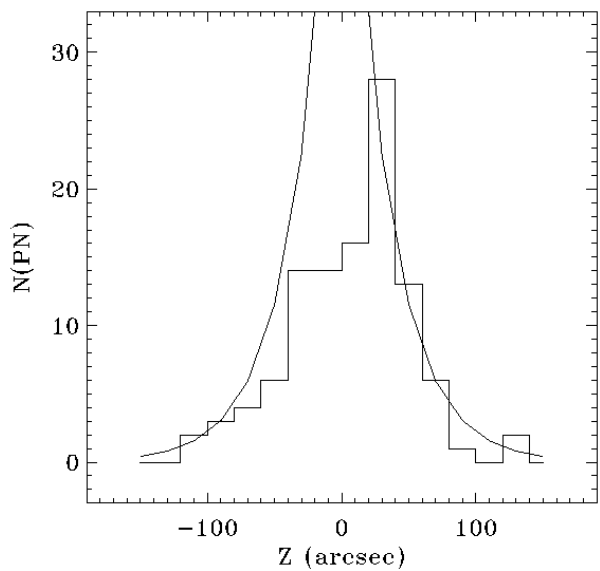}
\caption{Histogram showing number of PNe as a function of $z$ coordinate,
with 20 arcsec bins. The distribution is fitted with an exponential 
function having a scale height of 30 arcsec, 
equivalent to $\sim 500$ pc at the 
distance of M 82.
}
\label{scalhei}
\end{figure}

The presence of bright PNe at high $z$ is certainly not unique to M 82. 
For example, Ciardullo et al. (1991) found 33 PNe in the edge-on galaxy 
NGC 891, 21 of which were
more than 1 kpc distant from the galactic plane. These authors attributed
the high-$z$ PNe to a spheroidal component of NGC 891, presumably an old 
population, and used this as an argument favoring the universality of the
PNLF and its usefulness as a distance estimator.  Are the high-$z$ PNe
in M 82 also representatives of a stellar halo? Figure \ref{scalhei}
shows a histogram with the number of PNe as a function of $z$, with
20 arcsec bins. The distribution is roughly fitted if we assume an 
exponential function with a scale height of 500 pc. A similar exercise 
involving the 33 PNe in NGC 891 gives more or less the same result.

A rotating halo population is not the only possible interpretation.  As 
mentioned in \S \ 1, there is strong evidence for a recent tidal interaction 
between M 81 and M 82.  While none of our PNe lie coincident with any 
of these documented tidal features (Yun et al. 1994), nor do they show 
any outstanding peculiar motions, it can be argued that the 
disk of M 82 is very thick because it has been dynamically heated by 
the past interaction with M 81; so the high-$z$ PNe could be attributed 
to a disturbed, thick disk component.

Also, one could argue that because M 82 has been found to have an
extended disk, reaching out to $\sim 12$ kpc (Davidge 2008), these 
objects may in fact be members of the outer disk, found close to the 
galaxy midplane, seen in projection.  For instance, a $z$ distance 
of 55 arcsec translates to a position on the galaxy mid-plane 
$\sim 4$ kpc from the center, if it is projected straight
towards the observer.  In this situation, however, the conflict 
increases between the Sofue et al. (1992) rotation curve 
and our observed PN velocities, as objects rotating on the edge
nearest to the observer should have the smallest radial velocities, due 
to their high tangential velocity component.  Furthermore, 
PNe at these great radii should have extremely small rotational
speeds according to a falling rotation curve, thus making this
explanation of the high-$z$ rotation less attractive.

Of these three possibilities for the origin of the high rotational velocities
sustained by these high-$z$ objects, our current data is insufficient to 
provide any further elucidating interpretation.  One way to investigate this question 
further would be to obtain abundance measurements of these high-$z$ objects; 
a preponderance of low metallicities would confirm the stellar halo 
interpretation, and high metallicities would favor an extreme thick disk. 
Without such information, we cannot accurately make any further 
interpretation of our results and leave this as an open question for 
subsequent research.

\section{Photometry, PNLF, and Distance}

\subsection{[OIII] $\lambda$5007 Photometry}

In order to build our observed PNLF and obtain a distance estimate for 
M 82, we must first obtain [O{\textsc{iii}}] $\lambda$5007 magnitudes for 
each of our PN candidates.  These magnitudes are expressed using the 
standard definition for magnitudes $m(5007)$ from Jacoby (1989), 
\begin{equation}
m(5007)=-2.5\;log\;I(5007) - 13.74.
\end{equation}
We conducted photometry on our imaging data, analyzing each of 
the FOCAS CCD chips separately, due to slight photometric differences
that exist between the two chips.  Our measurements were calibrated
using observations of our spectrophotometric standard star, LTT 9491
(Colina \& Bohlin 1994).  Large aperture photometry using the
 IRAF \texttt{phot} task was conducted on single reference frames for
 our standard star and three bright, isolated, internal standard stars in 
 our target field.  This tied the spectrophotometric standard to our
images, and we continue from this point forward using differential 
photometry.

These internal frame standards were then measured using a small aperture on
the combined on-band frame, which allowed us to calculate an
aperture correction, as well as a correction that 
accounted for non-photometric conditions
introduced in the image stacking process.  From here, we switch
to PSF-fitting photometry using the DAOPHOT 
package (Stetson 1987) and its \texttt{psf} and
\texttt{allstar} tasks.  Using the internal frame standards, and four bright
PN candidates as additional standards, we measure PSF aperture
corrections and then perform PSF-fitting photometry on both the continuum
subtracted image and the box median subtracted image for all 109 PN candidates.

Two further photometric corrections were also performed.  
The first accounted for
a difference in airmass between the standard star image $(X=1.28)$ and our
reference image $(X=1.77)$ that affected the 
photometry at the point when we tied our
spectrophotometric standard to our internal frame standards.  This represents
only a small $-0.06$ mag correction to our measured, apparent magnitudes.  
The second
correction accounted for filter transmission so that total physical fluxes
from the emission lines could be properly measured.  Following a procedure
set out in Jacoby, Quigley, \& Africano (1987), we make our correction 
based on the filter transmission of the 
redshifted wavelength of the [O{\textsc{iii}}]
$\lambda$5007 line, which is 5010 \AA \  considering 
M 82's systemic velocity of
200 km s$^{-1}$.  Transmission at this wavelength 
is at the filter's peak value of 68\%,
and the corresponding magnitude correction for 
this transmission is $-0.419$ mag.

We then derived Jacoby magnitudes for all of our PN candidates.
Magnitudes were measured on both continuum subtracted and median images,
and preference was given to magnitudes derived from the continuum
subtracted images whenever these measurements were possible.  Magnitudes
derived from the median image were used when sources lay on 
unusable portions of the continuum subtracted image (masked regions that
contained bad pixels, saturated stars, regions of bright
background galaxy emission).  In all, 105 of 109 sources were successfully 
measured, and their $m(5007)$ magnitudes are listed in the source catalog
found in Table \ref{mytable}.

\begin{figure}
\epsscale{1.2}
\plotone{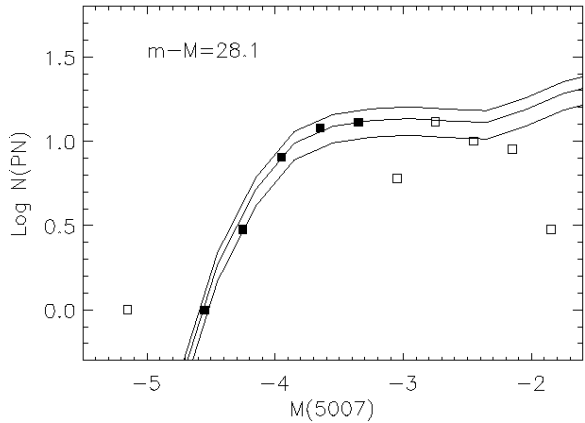}
\caption{Observed [O {\textsc{iii}}] $\lambda$5007 PNLF with 84 PNe binned
  at 0.3 mag intervals. Fits are made to the filled squares, while the open 
  squares on the right denote bins beyond our incompleteness threshold, and the open 
  square on the left denotes the bin for our single outlier object, PN 104.
  The absolute magnitudes are calculated using a best
  fit distance modulus of $m - M = 28.1$ and a foreground extinction correction of 
  0.28 mag.  The fitted theoretical PNLF is shown with sample sizes of
  810, 690, and 550 PNe (see M\'endez and Soffner 1997).}
\label{pnlf}
\end{figure}

\subsection{PNLF Construction and Fitting}

From the sources with measured Jacoby magnitudes, we would like to select a
statistically complete sample of PNe from which to build an accurate PNLF.
Objects chosen for this sample come from regions in the field of view where 
we are confident 
in our measurement of a complete population.  We want to eliminate PNe
detected in regions of high galaxy background emission where PNe above our
photometric incompleteness limit would still 
miss being detected because of their
inability to rise above high background flux levels.  
In order to accomplish this,
we lay down an isophote at the boundary where obvious, 
bright continuum emission becomes an obstacle
to our PN identifications, and eliminate objects from 
the statistical sample that 
fall within this region.  Also, we make a visual inspection of both our 
on-band and high-resolution H$\alpha$ images in order to reject objects
that lie on or near obvious dust features.  While defining these criteria 
remains a subjective practice, the total number of objects that contribute to 
our PNLF make our final result quite invariant to slight changes in this 
selection process.  

In the end, we select 84 PNe for our statistical sample.  From these 
selected [O{\textsc{iii}}] $\lambda$5007 magnitudes, we build our PNLF.  
To this distribution, we fit a theoretical PNLF curve 
generated as in M\'endez \& Soffner (1997). We could have used more
recent PNLF generation procedures, like in M\'endez et al. (2008), but
the end result at the bright end of the PNLF, which is what we use in 
PNLF distance determinations, is indistinguishable.
The simulated PNLF is fit to our distribution and plotted in 
Figure \ref{pnlf}.

There are many sources of uncertainty to consider when determining the final
factor of error for our distance determination.  Photometric error and
uncertainty in the PNLF fitting process each account for sources of random
error, at 0.1 mag each.  We also include 0.05 mag of uncertainty for filter
calibration into this factor of random error.  However, we must 
also consider systematic errors, such as uncertainty in the distance to
M 31, the calibrating galaxy for the PNLF distance measurement.  Following
Jacoby et al. (1990), these account for 0.13 mag of error in total.  
When added quadratically, random and systematic sources total to 
our final error estimate of 0.2 mag.

By combining our best fit PNLF with an adopted galactic extinction of 
$A_{5007}=0.28$ mag (see discussion below), our derived distance 
modulus is $28.1 \pm 0.2$, which is equivalent to a distance of 
$4.2 \pm 0.4$ Mpc.

\subsection{Galactic Foreground and Internal Extinction Effects}

\begin{figure}
\epsscale{1.15}
\plotone{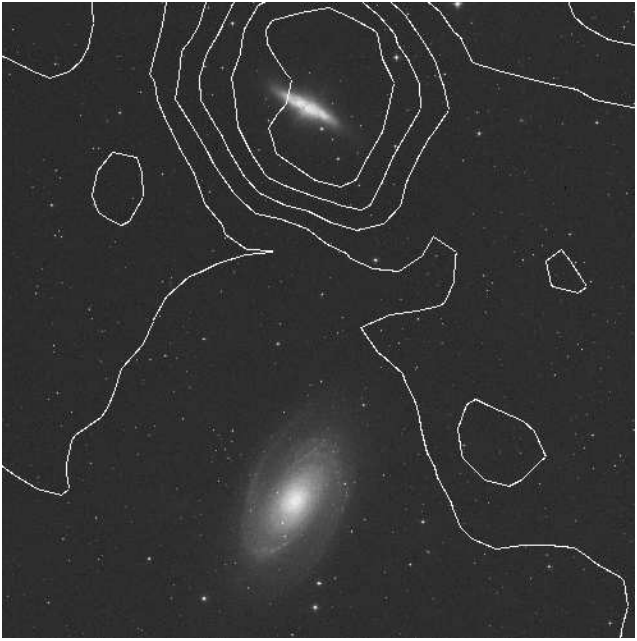}
\caption{Schlegel et al. (1998) dust map contours overlaid on DSS image of 
M 81/M 82 sky field.  Image is in standard 
orientation with north up, east left.}
\label{schlegel}
\end{figure}

One critical component of our distance determination lies in our adopted value
of galactic foreground extinction.  While Schlegel et al. (1998) has become 
a standard source for extinction values, we argue here that their value of
$A_V=0.526$ is incorrect.  Upon examining the dust maps from which the 
Schlegel et al. values are derived, it becomes immediately apparent that 
M 82's emission contribution was left unsubtracted from this galactic map
and our target is therefore assigned
an inappropriately high extinction value.  Figure \ref{schlegel} shows contours
created from the dust map overlaid atop an optical DSS image of the
nearby field of view, which includes the nearby galaxy M 81.  This figure 
clearly shows the point source nature of this emission.  

Under further scrutiny, the situation actually worsens.  The region of sky 
surrounding M 81 and M 82 has long been known to host troublesome 
highly spatially-variable foreground galactic extinction.  Sandage (1976) 
presents an image of bright, filamentary reflection nebulae covering
the entire field around these two galaxies.  With such highly variable
nearby structure, the reader should note that any single value
of foreground extinction is simply an estimation which carries a considerable
amount of uncertainty.  To carry out a robust foreground extinction correction,
individual estimates towards each PN should be made.  However, as the 
uncertainty of internal extinction dominates over the foreground component, 
we will continue to use a simple single-value approximation here.

In an effort to choose a proper foreground extinction estimation, we
exclude the value given by Burstein \& Heiles (1984) of $A_V=0.09$ as 
unreasonably low.  We instead adopt a value of $A_V=0.25$ from 
Dalcanton et al. (2009), who interpolated the value of extinction at M 82's
position from nearby, non-contaminated values in the Schlegel et al. (1998)
dust map.  For comparison, M 81 has a listed value of $A_V=0.26$.
Assuming a standard R=3.1 dust curve, this translates to our adopted
foreground extinction value of $A_{5007}=0.28$.

\begin{figure*}
\epsscale{1.1}
\plotone{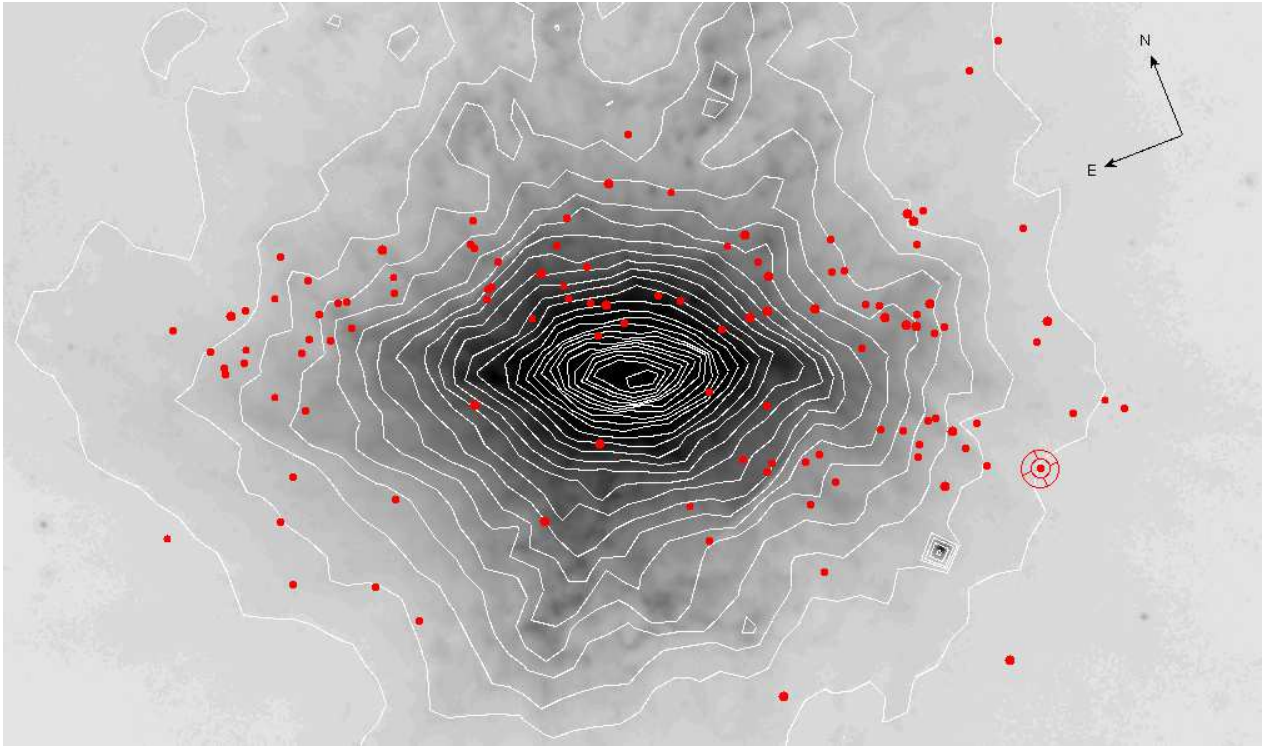}
\caption{Positions of the full 109-member PN sample overlaid on a SINGS IRAC 
8 $\mu$m image (Kennicutt 2003) showing hot dust emission in M 82.  The location of 
outlier PN 104 is denoted by the target symbol.}
\label{iracdust}
\end{figure*}

Our derived distance measurement of 4.2 $\pm 0.4$ Mpc is quite discrepant 
from the Dalcanton et al. (2009) TRGB-derived value for M 82 of 3.55 
$\pm 0.06$ Mpc .  We note that Sakai \& Madore (1999) also 
make a TRGB distance determination to M 82, however the Dalcanton et al. 
(2009) value takes precedence due to their superior data quality, star 
counts, and observational footprint away from obscuring dust.  In addition, 
we also derive a larger distance than those derived for group member M 81:
3.63 $\pm 0.34$ Mpc as calculated by Freedman et al. (1994) using Cepheid 
variables, and 3.50 $\pm 0.40$ as calculated by Jacoby et al. (1989) using
a similar PNLF fitting technique.  In order to reconcile this 
distance discrepancy, we point to internal 
extinction within M 82 as the most likely culprit.  As stated in \S \ 1, 
M 82 is a starbursting system with ongoing star formation;
it should be assumed to contain large amounts of obscuring
dust throughout the plane of the galaxy, into which we look edge-on.  
As evidence of this fact, Puxley (1991) estimates extinction as high as
$A_V=27$ to the center of the galaxy.  Engelbracht et al. (2006) studied 
a Spitzer IRAC image at 8$\mu$m where they showed that emission from 
hot dust was not only distributed throughout the plane, but also at heights 
up to 6 kpc from the plane.  Not only is the amount of dust 
plentiful in this system, but also distributed over a large extent, both in 
the plane and vertically.  To compare the positional distribution 
of our objects to the dust distribution surrounding M 82, we overlay the 
positions of our objects on this same 8$\mu$m image in Figure \ref{iracdust}.

One method to investigate how internal extinction is affecting our PNLF
is to study two sub-samples of the PNe: objects at greatest risk to internal 
extinction and those least susceptible to such effects.  In most cases, these 
sub-samples would show systematically different photometric results, and this
would help to constrain the effect of internal extinction on our results.  To attempt
this exercise, we first divide our sample according to minor axis offset, considering
that PNe that lie towards the plane have a greater likelihood of suffering from
internal extinction from dust that lies predominately in the plane.  Next, we
assign our sub-samples according to the amount of 8$\mu$m dust emission
detected at the positions of our PN objects, which takes into consideration the dust
that is known to lie vertically away from the plane.  Unfortunately, no detectable
systematic differences in the magnitude distributions were observed for 
sub-samples defined by either method.  The failure of this exercise may stem from
the fact that projection effects may cause a considerable variation of internal
extinction across small projected spatial scales, thus foiling any systematic search
for such a correlation.

\begin{figure}[b]
\epsscale{1.2}
\plotone{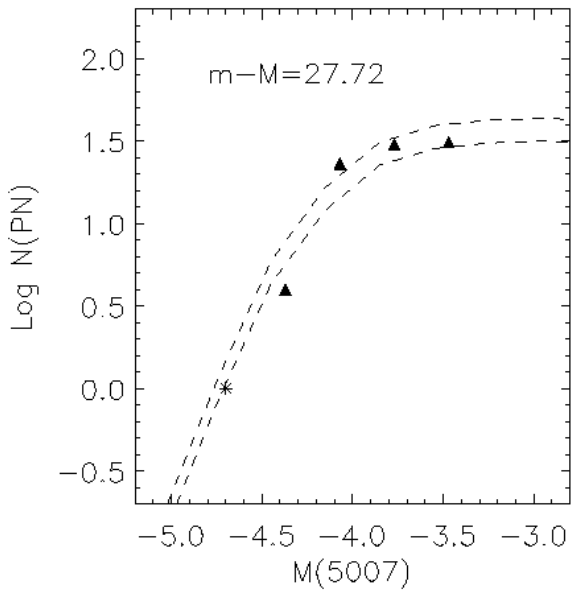}
\caption{Triangle symbols represent observed [O {\textsc{iii}}] $\lambda$5007 PNLF 
  for 88 PNe in M 81 from Jacoby et al. (1989) plotted with our simulated PNLF function
   to their best-fit distance modulus of $m - M = 27.72$ and a foreground extinction correction 
   of 0.36 mag.  The theoretical curves are shown for sample sizes of 2200 and 1600 PNe 
   (see M\'endez and Soffner 1997).  The asterisk symbol denotes the location of our 
   outlier PN 104 when considered at the same distance modulus and a foreground
   extinction correction of 0.28 mag.}
\label{m81pnlf}
\end{figure}

One observational signature that indeed suggests that internal extinction is 
responsible for our distance discrepancy appears as a lone outlier in our 
PNLF.  This single PN, ID = 104, appears $\sim 0.5$ magnitudes brighter
than the rest of the distribution.  While other PNLF studies 
suggest that these outliers might merely be persistent contaminants, it
is important to note that this PN candidate was throughly analyzed and
cleanly passes all five selection criteria.  We postulate that this object,
unlike the majority of the rest of our sample, does not suffer from any great
amount of internal extinction.  To support this claim, we point out that in
Figure \ref{iracdust}, this single object appears offset from the galaxy center
by $\sim 170$ arcsec in a region with little apparent dust in any of our 
images.  There are only four other PNe who appear at locations with less 
8$\mu$m emission, thus suggesting that this notion of an exception to our
otherwise extreme internal extinction is not unfounded.  Also, it is interesting to note 
that if this were indeed the single brightest PN in M 82 and we were to fit
our PNLF with a sensible sample size to this single data point, we would 
derive a $\sim 0.4$ mag lower best-fit distance modulus.  This new distance 
modulus of $27.7 \pm 0.2$ would agree well with the Dalcanton et al. (2009) 
distance modulus of $27.75 \pm 0.04$, along with many of the other comparable distance
measurements.  In particular, when assuming the same foreground
dust extinction estimate as used above, we can overlay the position of PN 104 to fit perfectly
along the best-fit PNLF derived by Jacoby et al. (1989) for M 81, as plotted 
in Figure \ref{m81pnlf}.

While the pieces fall into place well for the case that our data suffers from a
few tenths of a magnitude of internal extinction, the only true way to answer 
how much internal extinction affects our PNLF would be to acquire spectra 
for each of our PNe and measure individual extinctions to each object by means
of the Balmer decrement.  With these observations, we would be able to correct
for extinction on a point-by-point basis, instead of using broad statistical
corrections at a cost of increased uncertainties.

\section{Summary of Conclusions}

Using a method of slitless spectroscopy, we have
detected 109 PNe in M 82, measured their
radial velocities, and performed their photometry. We have carefully 
calibrated the radial velocity procedure
and we have tested it using observations of the local Galactic PN NGC 7293.
We find radial velocity uncertainties to be below 10 km s$^{-1}$. 

The average radial velocity of PNe close to the major axis in M 82 shows 
excellent agreement with previous velocity studies of 
the galaxy obtained by other methods 
(i.e. CO emission), further confirming our error estimates.
The individual radial velocities of PNe in M 82 show a clear rotational 
signature, again in excellent agreement with previous studies.  Through 
simulations, we find that our data also agrees with a near-Keplerian 
rotation curve, though this conclusion is not definitive.
Also, we discover that PNe who appear at high $z$ positions also show 
a clear rotational signature, which conflicts with an earlier published result
that claims there is little or no rotation away from the disk.  Additional 
observations are required in order to pin down whether these peculiar 
objects are part of a halo population or members of a
thick disk population.

Our derived PNLF shows several indications that our PN photometry
suffers from considerable internal extinction, on the order of 0.3 or 0.4 mag,
thus explaining our discrepant distance modulus of 
$28.1 \pm 0.2$ and its associated distance of 4.2 $\pm 0.4$ Mpc.  
Additional spectroscopy of these PNe would allow us to derive individual 
extinction measurements and improve upon our current measurement.\\

\acknowledgements
{This work was made possible by the NSF Astronomy REU 
program at the Institute for Astronomy, University of Hawaii, 
award AST-9987896.  RHM would like to acknowledge support 
by NSF grant 0307489.  The authors kindly acknowledge the
help provided by the Subaru staff, particularly the support
astronomers Youichi Ohyama, Takashi Hattori, and Kentaro Aoki.
Finally, the authors acknowledge the anonymous referee for
his/her careful reading and helpful suggestions.
}

{\it Facilities:} \facility{Subaru (FOCAS)}, \facility{HST (ACS)}

\clearpage

\begin{deluxetable*}{lrcrc}
\tablecaption{FOCAS observations and calibrations \label{tbl-1}}
\tablewidth{0pt}
\tablehead{
\colhead{FOCAS Field} & \colhead{Configuration} &
\colhead{FOCAS number} &
\colhead{Exp (s)} & \colhead{Airmass\tablenotemark{a}}}
\startdata
 M 82            &  Off-band  &  55749  &   60  &  1.88 \\
 M 82            &   On-band  &  55751  &  600  &  1.87 \\
 M 82            &  On+grism  &  55753  & 1000  &  1.82 \\
 Th-Ar + mask    &  On+grism  &  55757  &   10  &  1.76 \\
 Th-Ar + mask    &   On-band  &  55759  &    4  &  1.75 \\
 M 82            &  Off-band  &  55761  &   30  &  1.73 \\
 M 82            &   On-band  &  55763  &  300  &  1.73 \\
 M 82            &  On+grism  &  55765  &  500  &  1.71 \\
 NGC 7293 + mask &   On-band  &  55855  &  200  &  1.36 \\
 NGC 7293 + mask &  On+grism  &  55857  &  300  &  1.35 \\
 Th-Ar + mask    &  On+grism  &  55859  &    4  &  1.34 \\
 NGC 7293 + mask &   On-band  &  55869  &  200  &  1.32 \\
 NGC 7293 + mask &  On+grism  &  55871  &  300  &  1.32 \\
 Th-Ar + mask    &  On+grism  &  55875  &    4  &  1.32 \\
 LTT 9491        &   On-band  &  55877  &   10  &  1.28 \\ 
 LTT 9491        &   On-band  &  55879  &   20  &  1.28 \\ 
 LTT 9491        &   On-band  &  55881  &   10  &  1.28 \\ 
 LTT 9491        &   On-band  &  55883  &   20  &  1.28 \\ 
 M 82            &  Off-band  &  55963  &   60  &  1.78 \\
 M 82            &   On-band  &  55965  &  600  &  1.77 \\
 M 82            &  On+grism  &  55967  & 1000  &  1.73 \\
 M 82            &  Off-band  &  55975  &   60  &  1.67 \\
 M 82            &   On-band  &  55977  &  600  &  1.66 \\
 M 82            &  On+grism  &  55979  & 1000  &  1.64 \\
 NGC 7293 + mask &   On-band  &  56035  &  200  &  1.35 \\
 NGC 7293 + mask &  On+grism  &  56037  &  300  &  1.34 \\
 Th-Ar + mask    &  On+grism  &  56041  &    4  &  1.33 \\
 LTT 9491        &   On-band  &  56053  &   10  &  1.29 \\ 
 LTT 9491        &   On-band  &  56055  &   20  &  1.29 \\ 
 LTT 9491        &   On-band  &  56057  &   10  &  1.28 \\ 
 LTT 9491        &   On-band  &  56059  &   20  &  1.28 \\ 
 M 82            &  Off-band  &  56143  &   60  &  1.78 \\
 M 82            &   On-band  &  56145  &  600  &  1.77 \\
 M 82            &  On+grism  &  56147  & 1000  &  1.74 \\
 Th-Ar + mask    &  On+grism  &  56151  &   10  &  1.69 \\
 Th-Ar + mask    &   On-band  &  56155  &    4  &  1.68 \\
 M 82            &  Off-band  &  56157  &   60  &  1.67 \\
 M 82            &   On-band  &  56159  &  600  &  1.67 \\
 M 82            &  On+grism  &  56161  & 1000  &  1.64 \\
\enddata
\tablenotetext{a}{The airmass corresponding to the middle of each exposure.}
\end{deluxetable*}


\begin{deluxetable*}{c|rr|rrr|rrr|r|r|c}
\tablecaption{Rejected PN Candidates\label{mybadtable}}
\tablewidth{0pt}
\tabletypesize{\small}
\tablehead{
\colhead{ID} & \colhead{X} & \colhead{Z} & 
\multicolumn{3}{c}{$\alpha$} & \multicolumn{3}{c}{$\delta$} & 
\colhead{Helio. RV} & \colhead{\ } & \colhead{\ }\\
\colhead{Number} & \colhead{(arcsec)} & \colhead{(arcsec)} & 
\multicolumn{3}{c}{(J2000)} & \multicolumn{3}{c}{(J2000)} & 
\colhead{(km s$^{-1}$)} & \colhead{\textit{m}(5007)} & \colhead{Rejection}
}
\startdata
 224 & -55.065 &   37.755 & 09 & 55 & 39.374 & 69 & 41 & 02.12 &  165.16 & 26.691 & Extension\\
 227 & -55.007 &   23.652 & 09 & 55 & 40.472 & 69 & 40 & 49.34 &  173.72 & 25.062 & H$\alpha$\\
 124 & -57.562 &  -41.564 & 09 & 55 & 45.124 & 69 & 39 & 50.01 &  227.81 & 26.524 & Extension\\
 233 & -20.144 &   26.379 & 09 & 55 & 46.317 & 69 & 41 & 05.96 &  117.06 & 25.252 & Extension\\
 126 & -33.043 &  -10.824 & 09 & 55 & 46.966 & 69 & 40 & 27.94 &  174.89 & 23.043 & Ext. \& H$\alpha$\\
 127 & -67.446 & -132.141 & 09 & 55 & 50.400 & 69 & 38 & 24.36 & -251.38 & 25.864 & Velocity\\
 236 &   2.311 &   16.472 & 09 & 55 & 51.034 & 69 & 41 & 06.18 &  179.60 & 24.809 & Extension\\

\enddata
\tablecomments{The X and Z coordinates have their origin at 
$\alpha$= $9^{\rm h}55^{\rm m}51^{\rm s}.77$ 
$\delta$=$69^{\circ}40^{'}50''\!\!.22$, where 
X and Z are defined along the major and minor axes, respectively.}
\end{deluxetable*}

\clearpage

\LongTables
\begin{deluxetable*}{c|rr|rrr|rrr|c|c|c}
\tablecaption{Detected PN Candidates \label{mytable}}
\tablewidth{0pt}
\tabletypesize{\small}
\tablehead{ \colhead{ID} & \colhead{X} & \colhead{Z} & \multicolumn{3}{c}{$\alpha$} & \multicolumn{3}{c}{$\delta$} & \colhead{Helio. RV} & \colhead{\ } & \colhead{\ } \\
\colhead{Number} & \colhead{(arcsec)} & \colhead{(arcsec)} & \multicolumn{3}{c}{(J2000)} & \multicolumn{3}{c}{(J2000)} & \colhead{(km s$^{-1}$)} & \colhead{\textit{m}(5007)} & \colhead{Notes}
}
\startdata
201 & -145.526 &  136.956 & 09 & 55 & 16.227 & 69 & 41 & 54.71 & $\mathellipsis$ & 27.006 & S \\
101 & -202.709 &  -11.699 & 09 & 55 & 17.880 & 69 & 39 & 19.30 & $\mathellipsis$ & 24.744 & S \\
102 & -194.546 &   -8.568 & 09 & 55 & 18.996 & 69 & 39 & 25.25 & $\mathellipsis$ & 24.878 & S \\
202 & -133.872 &  124.036 & 09 & 55 & 19.192 & 69 & 41 & 47.88 & $\mathellipsis$ & 25.526 & S \\
203 & -157.579 &   61.217 & 09 & 55 & 19.918 & 69 & 40 & 41.65 &  62.54 & 26.836 & S \\
204 & -168.920 &   24.166 & 09 & 55 & 20.843 & 69 & 40 & 03.77 & 112.33 & 25.795 & S \\
103 & -181.553 &  -14.324 & 09 & 55 & 21.642 & 69 & 39 & 25.15 & $\mathellipsis$ & 25.655 & S \\
205 & -165.199 &   15.045 & 09 & 55 & 22.155 & 69 & 39 & 57.03 & $\mathellipsis$ & 24.272 & S \\
275 & -165.166 &   14.491 & 09 & 55 & 22.217 & 69 & 39 & 56.52 & $\mathellipsis$ & 25.131 & S \\
104 & -168.716 &  -36.910 & 09 & 55 & 25.566 & 69 & 39 & 09.89 & 143.52 & 23.301 & S \\
206 & -116.901 &   66.628 & 09 & 55 & 26.435 & 69 & 41 & 03.15 &  87.36 & 25.560 & S \\
207 & -112.919 &   62.136 & 09 & 55 & 27.482 & 69 & 41 & 00.67 & 264.42 & 26.442 & S \\
208 & -110.247 &   65.230 & 09 & 55 & 27.684 & 69 & 41 & 04.49 &  76.40 & 26.173 & S \\
209 & -114.516 &   52.567 & 09 & 55 & 27.929 & 69 & 40 & 51.40 &  45.76 & 24.711 & S \\
210 & -126.841 &   20.083 & 09 & 55 & 28.326 & 69 & 40 & 16.97 &  35.61 & 25.482 & S \\
211 & -120.499 &   29.271 & 09 & 55 & 28.694 & 69 & 40 & 27.92 &  84.49 & 24.933 & S \\
105 & -141.819 &  -19.293 & 09 & 55 & 28.789 & 69 & 39 & 36.46 &  83.94 & 25.606 & S \\
212 & -122.718 &   17.244 & 09 & 55 & 29.239 & 69 & 40 & 16.10 &  46.05 & 25.166 & S \\
106 & -146.651 &  -36.563 & 09 & 55 & 29.304 & 69 & 39 & 18.94 & 147.85 & 24.602 & S \\
213 & -115.346 &   24.809 & 09 & 55 & 29.940 & 69 & 40 & 25.88 & 122.30 & 24.789 & S \\
214 & -115.254 &   19.696 & 09 & 55 & 30.338 & 69 & 40 & 21.33 & -17.54 & 25.417 & S \\
107 & -137.401 &  -29.834 & 09 & 55 & 30.354 & 69 & 39 & 28.68 &  36.14 & 25.785 & S \\
108 & -131.814 &  -23.061 & 09 & 55 & 30.776 & 69 & 39 & 37.02 &  56.92 & 26.233 & \\
215 & -111.224 &   20.299 & 09 & 55 & 30.994 & 69 & 40 & 23.46 &  69.99 & 25.729 & S \\
109 & -125.155 &  -18.145 & 09 & 55 & 31.556 & 69 & 39 & 44.12 & 192.32 & 25.514 & \\
110 & -122.160 &  -18.962 & 09 & 55 & 32.142 & 69 & 39 & 44.61 & 143.14 & 25.850 & \\
216 & -102.429 &   22.912 & 09 & 55 & 32.302 & 69 & 40 & 29.42 &  94.00 & 24.485 & S \\
217 & -100.137 &   27.390 & 09 & 55 & 32.361 & 69 & 40 & 34.45 &  56.40 & 23.749 & S \\
111 & -129.677 &  -45.397 & 09 & 55 & 32.904 & 69 & 39 & 17.71 &  61.16 & 24.885 & S \\
218 &  -94.876 &   27.588 & 09 & 55 & 33.249 & 69 & 40 & 36.76 & 115.83 & 25.612 & S \\
112 & -159.255 & -115.145 & 09 & 55 & 33.304 & 69 & 38 & 03.50 & $\mathellipsis$ & 25.909 & S \\
113 & -118.861 &  -28.890 & 09 & 55 & 33.462 & 69 & 39 & 36.85 &  35.52 & 24.330 & S \\
219 &  -85.699 &   41.002 & 09 & 55 & 33.806 & 69 & 40 & 52.67 &  58.71 & 26.105 & S \\
220 &  -79.817 &   53.573 & 09 & 55 & 33.857 & 69 & 41 & 06.43 &  95.05 & 25.136 & S \\
114 & -118.427 &  -33.846 & 09 & 55 & 33.928 & 69 & 39 & 32.58 &  42.83 & 25.230 & S \\
115 & -111.888 &  -23.672 & 09 & 55 & 34.278 & 69 & 39 & 44.45 &  42.71 & 26.143 & S \\
221 &  -80.695 &   40.060 & 09 & 55 & 34.728 & 69 & 40 & 53.80 &  24.91 & 25.039 & S \\
222 &  -93.733 &    9.993 & 09 & 55 & 34.801 & 69 & 40 & 21.30 & $\mathellipsis$ & $\mathellipsis$ & \\
116 & -102.588 &  -23.283 & 09 & 55 & 35.832 & 69 & 39 & 48.54 &  80.61 & 24.618 & S \\
223 & -74.015 &   25.360 & 09 & 55 & 37.031 & 69 & 40 & 43.11 &   74.90 & 24.727 & S \\
225 & -50.682 &   43.496 & 09 & 55 & 39.692 & 69 & 41 & 09.22 &  124.81 & 24.751 & S \\
226 & -44.966 &   53.954 & 09 & 55 & 39.885 & 69 & 41 & 20.92 &  231.09 & 26.354 & S \\
117 & -85.562 &  -44.888 & 09 & 55 & 40.481 & 69 & 39 & 35.78 &   71.57 & 25.802 & S \\
118 & -78.220 &  -34.265 & 09 & 55 & 40.930 & 69 & 39 & 48.32 &   24.03 & 26.249 & S \\
228 & -38.376 &   48.845 & 09 & 55 & 41.424 & 69 & 41 & 19.00 &  139.57 & 24.485 & S \\
229 & -48.021 &   20.959 & 09 & 55 & 41.904 & 69 & 40 & 49.65 &   99.32 & 25.967 & \\
119 & -72.892 &  -37.564 & 09 & 55 & 42.108 & 69 & 39 & 47.51 &  138.35 & 25.889 & S \\
120 & -75.375 &  -54.511 & 09 & 55 & 43.004 & 69 & 39 & 31.18 &  159.16 & 26.207 & S \\
121 & -56.824 &  -15.421 & 09 & 55 & 43.187 & 69 & 40 & 14.24 &  157.94 & 25.845 & \\
230 & -14.902 &   69.494 & 09 & 55 & 43.873 & 69 & 41 & 47.32 &  234.07 & 26.115 & S \\
122 & -82.030 &  -81.482 & 09 & 55 & 43.952 & 69 & 39 & 04.10 &   67.92 & 24.960 & S \\
231 & -37.324 &   15.586 & 09 & 55 & 44.191 & 69 & 40 & 49.17 &  133.61 & 24.181 & \\
123 & -59.248 &  -38.159 & 09 & 55 & 44.533 & 69 & 39 & 52.54 &  203.52 & 27.042 & S \\
232 &   3.495 &   92.526 & 09 & 55 & 45.331 & 69 & 42 & 15.56 &  159.98 & 26.462 & S \\
125 & -47.574 &  -37.199 & 09 & 55 & 46.491 & 69 & 39 & 58.03 &  201.88 & 25.009 & S \\
234 & -11.200 &   28.132 & 09 & 55 & 47.770 & 69 & 41 & 11.37 &  203.90 & 25.508 & \\
235 &  10.607 &   72.570 & 09 & 55 & 48.120 & 69 & 42 & 00.43 &  196.41 & 25.265 & S \\
128 & -35.211 &  -70.671 & 09 & 55 & 51.228 & 69 & 39 & 32.76 &   56.54 & 25.927 & S \\
129 & -27.275 &  -57.126 & 09 & 55 & 51.579 & 69 & 39 & 48.17 &  185.94 & 26.948 & S \\
237 &   9.864 &   23.792 & 09 & 55 & 51.782 & 69 & 41 & 15.83 &  176.03 & 25.370 & \\
238 &  26.476 &   57.809 & 09 & 55 & 52.048 & 69 & 41 & 53.50 &  269.85 & 26.180 & S \\
239 &  18.273 &   38.672 & 09 & 55 & 52.084 & 69 & 41 & 32.79 &  224.30 & 26.630 & S \\
240 &  15.678 &   23.991 & 09 & 55 & 52.788 & 69 & 41 & 18.47 &  290.38 & 25.684 & \\
241 &  12.693 &   11.284 & 09 & 55 & 53.259 & 69 & 41 & 05.66 &  178.72 & 25.234 & \\
242 &  30.448 &   46.753 & 09 & 55 & 53.618 & 69 & 41 & 45.15 &  165.07 & 26.027 & S \\
243 &  24.555 &   25.568 & 09 & 55 & 54.211 & 69 & 41 & 23.44 &  270.29 & 25.981 & \\
244 &  26.764 &   30.331 & 09 & 55 & 54.229 & 69 & 41 & 28.69 &  261.99 & 26.008 & S \\
245 &  36.476 &   35.505 & 09 & 55 & 55.522 & 69 & 41 & 37.25 &  220.12 & 25.560 & S \\
130 &  10.011 &  -32.892 & 09 & 55 & 56.201 & 69 & 40 & 25.37 &  370.60 & 25.739 & \\
246 &  39.033 &   17.096 & 09 & 55 & 57.387 & 69 & 41 & 21.45 &  304.91 & 25.346 & \\
247 &  53.594 &   39.257 & 09 & 55 & 58.219 & 69 & 41 & 47.56 &  264.42 & $\mathellipsis$ & \\
248 &  64.343 &   55.899 & 09 & 55 & 58.802 & 69 & 42 & 06.97 &  232.40 & $\mathellipsis$ & \\    
249 &  55.625 &   29.140 & 09 & 55 & 59.356 & 69 & 41 & 39.26 &  232.55 & 25.054 & \\
250 &  63.310 &   44.899 & 09 & 55 & 59.479 & 69 & 41 & 56.64 &  220.82 & 24.849 & S \\
251 &  64.794 &   46.037 & 09 & 55 & 59.655 & 69 & 41 & 58.28 &  215.06 & 25.634 & S \\
252 &  57.287 &   28.390 & 09 & 55 & 59.714 & 69 & 41 & 39.22 &  223.46 & 24.867 & \\
253 &  57.531 &   24.198 & 09 & 56 & 00.075 & 69 & 41 & 35.50 &  196.44 & 25.851 & \\
131 &  31.081 &  -64.879 & 09 & 56 & 02.358 & 69 & 40 & 04.81 &   93.79 & 26.178 & S \\
132 &  60.722 &  -19.241 & 09 & 56 & 04.005 & 69 & 40 & 58.36 &  340.08 & 23.840 & \\
254 &  95.367 &   31.718 & 09 & 56 & 06.073 & 69 & 41 & 57.57 &  270.73 & 24.670 & S \\
255 & 100.668 &   42.853 & 09 & 56 & 06.133 & 69 & 42 & 09.73 &  298.15 & 24.260 & S \\
256 &  95.192 &   25.388 & 09 & 56 & 06.527 & 69 & 41 & 51.71 &  305.05 & 25.034 & S \\
257 & 114.435 &   21.569 & 09 & 56 & 10.153 & 69 & 41 & 55.94 &  252.49 & 25.112 & S \\
258 & 112.151 &   11.089 & 09 & 56 & 10.539 & 69 & 41 & 45.48 &  319.64 & 24.436 & \\
259 & 117.595 &   20.688 & 09 & 56 & 10.793 & 69 & 41 & 56.46 &  369.32 & 24.794 & S \\
133 &  91.250 &  -58.375 & 09 & 56 & 12.299 & 69 & 40 & 35.07 &  171.79 & 25.404 & S \\
260 & 130.628 &   29.174 & 09 & 56 & 12.359 & 69 & 42 & 09.24 &  126.48 & 25.025 & S \\
261 & 120.283 &    5.840 & 09 & 56 & 12.404 & 69 & 41 & 43.97 &  392.85 & 25.563 & \\
262 & 125.468 &   15.685 & 09 & 56 & 12.530 & 69 & 41 & 54.99 &  297.28 & 24.508 & S \\
263 & 141.626 &   38.460 & 09 & 56 & 13.553 & 69 & 42 & 22.06 &  186.35 & 25.309 & S \\
264 & 128.699 &    5.624 & 09 & 56 & 13.884 & 69 & 41 & 47.25 &  352.74 & 24.389 & \\
134 &  80.541 & -107.137 & 09 & 56 & 14.172 & 69 & 39 & 46.60 &  198.22 & 24.896 & S \\
265 & 132.073 &   -0.017 & 09 & 56 & 14.877 & 69 & 41 & 43.32 &  302.95 & 25.073 & \\
266 & 143.288 &   21.424 & 09 & 56 & 15.162 & 69 & 42 & 07.28 &  313.53 & 25.594 & S \\
135 & 129.293 &  -24.114 & 09 & 56 & 16.221 & 69 & 41 & 21.41 &  232.96 & 24.350 & S \\
136 &  98.689 &  -94.201 & 09 & 56 & 16.312 & 69 & 40 & 05.75 &  288.07 & 25.912 & S \\
267 & 155.104 &   16.468 & 09 & 56 & 17.592 & 69 & 42 & 07.52 &  388.14 & 26.875 & S \\
137 & 141.628 &  -19.384 & 09 & 56 & 17.999 & 69 & 41 & 30.70 & $\mathellipsis$ & 24.072 & S \\
268 & 154.513 &    0.252 & 09 & 56 & 18.688 & 69 & 41 & 52.55 &  252.46 & 24.299 & S \\
269 & 161.506 &   14.006 & 09 & 56 & 18.839 & 69 & 42 & 07.80 &  344.87 & 25.516 & S \\
138 & 133.244 &  -51.032 & 09 & 56 & 18.969 & 69 & 40 & 58.71 &  284.32 & 24.820 & S \\
270 & 155.134 &   -4.830 & 09 & 56 & 19.193 & 69 & 41 & 48.22 & $\mathellipsis$ & 24.416 & S \\
271 & 163.421 &   -7.060 & 09 & 56 & 20.801 & 69 & 41 & 49.57 & $\mathellipsis$ & 25.245 & S \\
272 & 162.508 &   -9.450 & 09 & 56 & 20.827 & 69 & 41 & 46.92 & $\mathellipsis$ & $\mathellipsis$ & \\
139 & 138.071 &  -69.051 & 09 & 56 & 21.182 & 69 & 40 & 44.21 &  261.26 & 25.842 & S \\
273 & 169.387 &   -0.872 & 09 & 56 & 21.341 & 69 & 41 & 57.47 &  319.03 & 24.382 & S \\
140 & 132.240 &  -94.696 & 09 & 56 & 22.111 & 69 & 40 & 18.79 &  336.34 & 25.776 & S \\
274 & 185.345 &    7.267 & 09 & 56 & 23.405 & 69 & 42 & 10.99 & $\mathellipsis$ & 24.954 & S \\
141 & 184.188 &  -78.470 & 09 & 56 & 29.704 & 69 & 40 & 54.19 & $\mathellipsis$ & 25.645 & S \\
\enddata
\tablecomments{The X and Z coordinates have their origin at 
$\alpha$= $9^{\rm h}55^{\rm m}51^{\rm s}.77$ 
$\delta$=$69^{\circ}40^{'}50''\!\!.22$, where 
X and Z are defined along the major and minor axes, 
respectively. An ``S'' denotes
membership in the statistical sample used for PNLF fitting.}
\end{deluxetable*}


\clearpage

\begin{thebibliography}{}

\bibitem[]{1} Achtermann, J.M., \& Lacy, J.H. 1995, ApJ, 439, 163

\bibitem[]{2} Acker, A. et al. 1992, Strasbourg-ESO Catalogue of Galactic Planetary Nebulae (Garching:ESO)

\bibitem[]{3} 
Alard, C., \& Lupton, R.H. 1998, ApJ, 503, 325

\bibitem[]{4} Arnaboldi, M., et al. 2008, ApJ, 674, L17

\bibitem[]{5}
Bottema, R. 1993,A\&A, 275,16

\bibitem[]{6}
Burstein, D., \& Heiles, C. 1984, ApJS, 54, 33

\bibitem[]{7}
Ciardullo, R., et al. 2002, ApJ, 577, 31

\bibitem[]{8}
Ciardullo, R. 2003, in IAU Symp. 209, Planetary Nebulae: 
Their Evolution and Role in the Universe, ed. S. Kwok, M. 
Dopita \& R. Sutherland (San Francisco: ASP), 617

\bibitem[]{9}
Ciardullo, R., et al. 1991, ApJ, 383, 487

\bibitem[]{10}
Colina, L., \& Bohlin, R.C. 1994, AJ, 108, 1931

\bibitem[]{11}
Cottrell, G. A.  1977, MNRAS, 178, 577

\bibitem[]{12}
de Grijs, R., et al. 2000, AJ, 119, 681

\bibitem[]{13}
Dalcanton, J., et al.  2009, ApJS, submitted

\bibitem[]{14}
Davidge, T.J.  2008, AJ, 136, 2502

\bibitem[]{15}
De Lorenzi, F., et al. 2008, MNRAS, 385, 1729

\bibitem[]{16}
Engelbracht, C.W., et al. 2006, ApJ, 642, L127

\bibitem[]{17}
Feldmeier, J.J., et al. 2004, ApJ, 615, 196

\bibitem[]{18}
Freedman, W.L., et al. 1994, ApJ, 427, 628

\bibitem[]{19}
G\"ossl, C.A., \& Riffeser, A. 2002, A\&A, 381, 1095

\bibitem[]{20}
Herrmann, K.A., et al.  2008, ApJ, 683, 630

\bibitem[]{21}
Hui, X., Ford, H.C., Freeman, K.C. \& Dopita, M.A. 1995, ApJ, 449, 592

\bibitem[]{22}
Jacoby, G.H., Ciardullo, R., \& Ford, H.C. 1990, ApJ, 356, 332

\bibitem[]{23}
Jacoby, G.H., Ciardullo, R., Ford, H.C., \& Booth, J. 1989, ApJ, 344, 704

\bibitem[]{24}
Jacoby, G.H., Quigley, R.J., \& Africano, J.L. 1987, PASP, 99, 672

\bibitem[]{25}
Kashikawa, N., et al. 2002, PASJ, 54, 819

\bibitem[]{26}
Kennicutt, R.C., et al. 2003, PASP, 115, 928

\bibitem[]{27}
Mayya, Y.D., Carrasco, L., \& Luna, A. 2005, ApJ, 628, L33

\bibitem[]{28}
McKeith, C.D., et al. 1993, A\&A, 272, 98

\bibitem[]{29}
Meaburn, J., et al. 2005, MNRAS, 360, 963

\bibitem[]{30}
M\'endez, R.H., \& Soffner, T. 1997, A\&A, 321, 898 

\bibitem[]{31}
M\'endez, R.H., et al. 2008, ApJ, 681, 325 

\bibitem[]{32}
M\'endez, R.H., et al. 2009, ApJ, 691, 228

\bibitem[]{33}
Mutchler, M., et al. 2007, PASP, 119, 1

\bibitem[]{34}
Napolitano et al. 2009, MNRAS, 393, 329

\bibitem[]{35}
Puxley, P.J. 1991, MNRAS, 249, 11

\bibitem[]{36}
Rodriguez-Rico, C. A., et al. 2004, ApJ, 616, 783

\bibitem[]{37}
Sakai, S., \& Madore, B.F. 1999, ApJ, 526, 599

\bibitem[]{38}
Sandage, A. 1976, AJ, 81, 954

\bibitem[]{39}
Schlegel, D.J., Finkbeiner, D.P., \& Davis, M. 1998, ApJ, 500, 525

\bibitem[]{40}
Shaver, P.A., et al. 1983, MNRAS, 204, 53

\bibitem[]{41}
Sofue, Y. 1998, PASJ, 50, 227

\bibitem[]{42}
Sofue, Y., et al. 1992, ApJ, 395, 126

\bibitem[]{43}
Stetson, P.B. 1987, PASP, 99, 191

\bibitem[]{44}
Telesco, C.M., et al. 1991, ApJ, 369, 135

\bibitem[]{45}
Yun, M. S. 1999, in IAU Symp. 186, Galaxy Interactions 
at Low and High Redshift, ed. J. E. Barnes \& D. B. 
Sanders (Dordrecht: Kluwer), 81

\bibitem[]{46}
Yun, M.S., Ho, P.T.P., \& Lo, K.Y. 1993, ApJ, 411, L17

\bibitem[]{47}
Yun, M.S., Ho, P.T.P., \& Lo, K.Y. 1994, Nature, 372, 530

\end{thebibliography}
\end{document}